\documentclass[superscriptaddress,twocolumn,amsmath,amssymb,floatfix,aps,prd,nofootinbib,notitlepage,preprintnumbers]{revtex4-2}
\usepackage[colorlinks = true,
            linkcolor = blue,
            urlcolor  = blue,
            citecolor = blue,
            anchorcolor = blue]{hyperref}
\usepackage{aas_macros}
\usepackage{amsmath,amssymb,amsfonts}
\usepackage{cancel}
\usepackage[dvipdfmx]{graphicx}
\usepackage{enumerate} 
\usepackage{colordvi} 
\usepackage{color} 
\usepackage{bm}

\newcommand{\mass}{m_\phi}
\newcommand{\mhalo}{M_{\rm h}}
\newcommand{\RR}{\mathcal{R}}
\newcommand{\rhoc}{\rho_{\rm c}}
\newcommand{\rhos}{\rho_{\rm s}}
\newcommand{\rs}{r_{\rm s}}
\newcommand{\rc}{r_{\rm c}}
\newcommand{\xc}{x_{\rm c}}
\newcommand{\cvir}{c_{\rm vir}}

\begin{document}
\preprint{YITP-22-82}

\title{Analytical approach to core-halo structure of fuzzy dark matter}
\author{Atsushi Taruya} \email{ataruya@yukawa.kyoto-u.ac.jp}
\affiliation{Center for Gravitational Physics and Quantum Information, Yukawa Institute for Theoretical Physics, Kyoto University, Kyoto 606-8502, Japan}
\affiliation{Kavli Institute for the Physics and Mathematics of the Universe (WPI), The University of Tokyo Institutes for Advanced Study, The University of Tokyo, 5-1-5 Kashiwanoha, Kashiwa, Chiba 277-8583, Japan}

\author{Shohei Saga} \email{saga@iap.fr}
\affiliation{Sorbonne Universit\'e, CNRS, UMR7095, Institut d'Astrophysique de Paris, 98bis boulevard Arago, F-75014 Paris, France}
\affiliation{Laboratoire Univers et Th{\'e}ories, Observatoire de Paris, Universit{\'e} PSL, CNRS, Universit{\'e} de Paris, 5 place Jules Janssen 92190 Meudon, France}
\date{\today}
\begin{abstract}
Ultralight bosonic dark matter called fuzzy dark matter (FDM) has attracted much attention as an alternative to the cold dark matter. An intriguing feature of the FDM model is the presence of a soliton core, a stable dense core formed at the center of halos. In this paper, we analytically study the dependence of the soliton core properties on the halo characteristics by solving approximately the Schr\"odinger-Poisson equation. Focusing on the ground-state eigenfunction, we derive a key expression for the soliton core radius, from which we obtain the core-halo mass relations similar to those found in the early numerical work, but involving the factor dependent crucially on the halo concentration and cosmological parameters. Based on the new relations, we find that for a given cosmology, (i) there exist a theoretical bound on the radius and mass of soliton core for each halo mass (ii) incorporating the concentration-halo mass (C-M) relation into the predictions, the core-halo relations generally exhibit a non-power-law behavior, and with the C-M relation suppressed at the low-mass scales, relevant to the FDM model, predictions tend to match the simulations well (iii) the scatter in the C-M relation produces a sizable dispersion in the core-halo relations, and can explain the results obtained from cosmological simulations. Finally, the validity of our analytical treatment are critically examined. A perturbative estimation suggests that the prediction of the core-halo relations is valid over a wide range of parameter space, and the impact of the approximation invoked in the analytical calculations is small, although it is not entirely negligible. 
\end{abstract}

\maketitle

\section{Introduction}
\label{sec:introduction}

The dark matter, invisible matter component, is an essential constituent accounting for $\sim30\%$ of the energy contents of the Universe. Although the origin of dark matter is not yet clarified, the structure formation driven by the gravitational instability, consistent with large-scale structure observations, suggests that the dark matter is nonbaryonic, and have negligible velocity dispersion, referred to as the cold dark matter (CDM). Since its foundation in early 1980s \cite{Peebles1982,Bond_Szalay_Tuerner1982,Blumenthal_etal1984}, the CDM has been playing an important role to establish a concordant picture of the Universe, and it successfully describes a wide range of observations on scales larger than galaxies. However, on the subgalactic scales, recent observations suggest that deviation from CDM predictions is visible, and remains significant especially for the observations of the local Universe, highlighted as the small-scale crises (see e.g., \cite{Bullock_Boylan-Kolchin_review2017,Tulin_Yu_review2018} for review). 

To relieve possible flaws in the CDM paradigm, ultralight particle of masses, $m_\phi\sim10^{-22}$-$10^{-20}$eV, called fuzzy dark matter (FDM) \cite{Hu_Barkana_Gruzinov2000}, has recently attracted much attention as an alternative to the CDM (Refs.~\cite{Marsh_review2016,Niemeyer_review2020,Hui_review2021,Ferreira_review2020} for review, see also Ref.~\cite{Hui_Ostriker_Tremaine_Witten2017}). While the predictions for the evolution and formation of large-scale structure remain essentially the same as in CDM model, and thus consistent with large-scale observations (e.g., Ref.~\cite{Hlozek_etal2018,Lague_etal2022} for recent works), many prominent features appear at small scales due to a very long de Broglie wavelength that extends over astrophysical scales, $\lambda_{\rm dB}=h/(\mass v)\simeq0.48\,{\rm kpc}\,(10^{-22}\,{\rm eV}/\mass)(250{\rm km}/v)$, with $v$ being the particle velocity. The FDM model can therefore modify small-scale structure formation, and thus the small-scale observations can place a tighter bound on the FDM model (e.g., Refs.~\cite{Armengaud_etal2017,Kobayashi_etal2017,Rogers_Peiris2021} for constraints from the Lyman-$\alpha$ forest). Importantly, modifications of the small-scale structures appears manifest not only through the small-scale cutoff of the linear power spectrum \cite{Hu_Barkana_Gruzinov2000} (see also Refs.~\cite{Khlopov_Malomed_Zeldovich1985,Nambu_Sasaki1990} for early works), but also by the quantum nature of dark matter particles at a macroscopic level. The latter aspect is particularly interesting, and can bring unique signature of FDM models to observationally test and constrain their model parameters.

Numerical studies on the FDM structure formation show that the dark matter halos commonly host a dense flat core called {\it soliton} with a universal density profile at the central region, accompanied by granular structures with a typical size of the de Broglie wavelength that extends over the outskirt of halos \cite{Schive_Chiueh_Broadhurst2014}. A notable feature of the soliton was its relation to the host halos. In early works, the mass and radius of FDM solitons, $M_{\rm c}$ and $\rc$, are found to be correlated tightly with halo masses in a power-law form as $M_{\rm c}\propto M_{\rm h}^\alpha$ and $\rc\propto M_{\rm h}^\beta$, and  Ref.~\cite{Shive_etal2014} obtained $\alpha\sim1/3$ and $\beta\sim-1/3$. Later, these properties have been examined with different setups by several groups \cite{Schwabe_etal2016,Du_etal2017,Mocz_etal2017,Mina_Mota_Winther2020}, and  obtained a similar result but with slightly different slopes (see also Ref.~\cite{Nori_Baldi2021} for  systematic study based on a particle-based method). More recent simulations covering a large cosmological volume suggest that not tight scaling relation exists between soliton structure and halo mass, and rather indicate a large scatter in the correlation \cite{May_Springel2021,Jowett_etal2022}. Another intriguing property of the soliton core comes from the results of high-resolution simulations \cite{Veltmaat_Niemeyer_Schwabe2018,Schive_Chiueh_Broadhurst2020}, showing that the soliton core is not strictly stable, but rather moves around a halo center in a random fashion, accompanying a rapid oscillation with the order-of-unity variation of its amplitude. These phenomena are considered as a consequence of the wave interference \cite{Li_Hui_Yavetz2021, Luna_etal2022}.

So far, most of these properties has been investigated by numerical simulations of the Schr\"odinger-Poisson equation, starting with various setups, and little analytical work has been done (but see \cite{Chavanis2019,Desjacques_Nusser2019}). While the soliton core-halo mass relation has led to a number of works that have tried to test and constrain the FDM model (e.g., Refs.~\cite{Safarzadeh_Spergel2020,Hayashi_Obata2020,Broadhurst_etal2020,Burkert2020,Hayashi_etal2021}), a major difficulty in characterizing such a relation in numerical simulations is a severe requirement of the dynamic range. That is, a realistic simulation must describe, in a relevant cosmological setup, both the halo merger processes and wave nature of soliton dynamics, typical scales of that, respectively, appear above Mpc and below kpc, thus requiring both a large simulation box and a sufficient spatial resolution. Even the state-of-the-art simulation is still challenging to trace an entire history of the formation and evolution of cosmic structures while fulfilling both the conditions. Aim of this paper is therefore to investigate analytically the soliton core structure, and to address the core-halo relations in a cosmological setup.

In doing so, a crucial step is how to solve the Schr\"odinger-Poisson equation. Coupled with the Poisson equation sourced by the mass density of FDM, the governing equation for the wave function of FDM becomes nonlinear. In particular, if one wants to address the soliton core structure, the contributions of both the soliton and surrounding halo structures have to be taken into account consistently. As we will describe in detail, for an analytically tractable calculation, our approach is to consider the contribution of halo as a smooth background, and to describe the soliton state ignoring its self gravity. Although the flat core structure is realized under a balance between quantum pressure and gravity and one thus expects the self-gravity of the soliton to play a certain role, this setup greatly simplifies the situation, and the problem is reduced to a linear eigenvalue problem. Still, there exists no exact solution, and some approximations need to be invoked to construct analytically the eigenfunctions of the soliton state. In this paper, we apply the uniform asymptotic approximation invented by Langer \cite{Langer1932,Langer1959,Olver1975} to the description of soliton state. In the presence of a large system parameter, this approximation is known to significantly outperform the familiar Wenzel-Kramers-Brillouin (WKB) approximation, and provides a very accurate way to estimate both eigenvalues and eigenfunctions. In particular, the analytically constructed eigenfunctions by this method is globally valid even around the so-called turning points, where the WKB approximation is broken down. Making use of the Langer's method, in this paper, we present analytical solutions of the Schr\"odinger-Poisson equation under a spherically symmetric halo potential. Focusing particularly on the ground-state wave function for the zero angular momentum $(\ell=0)$, we derive the analytical expressions for the core-halo relations. The resultant expressions apparently resemble those found in numerical simulations, but importantly, they include an additional factor dependent on cosmology and halo properties, with which the predicted core-halo relations can differ from a power-law form found in small-box simulations. We will discuss it in detail in comparison with simulation results. Further, we estimate the impact of the soliton self-gravity which we ignore on the core-halo relations, and the validity of our analytical treatment is discussed.

This paper is organized as follows. In Sec.~\ref{sec:basic_eq}, we begin by describing the setup of the problem to characterize the soliton core structure in a halo. Section~\ref{sec:Analytic_solutions_ell0} then presents the analytical construction of the soliton eigenstate, focusing on the cases with zero angular momentum. The construction for the nonzero angular momentum is presented in Appendix \ref{sec:Analytic_solutions_non_zero_ell}. Section~\ref{sec:result} presents the main results of this paper, and based on the treatment in previous section, we derive the analytical expressions of core-halo relations, and provide a general prediction for the core radius vs halo mass and core mass vs halo mass relations, treating the halo concentration as a free parameter. Further, imposing models of the concentration-halo mass relation (C-M relation), the predicted relations are compared with simulations. In Sec.~\ref{sec:discussion}, the validity of our treatment ignoring the soliton self-gravity is critically examined, and employing a perturbative calculation, we present an estimate of its impact on the soliton core radius. Finally, Sec.~\ref{sec:conclusion} is devoted to summary of important findings and conclusion. 

Throughout the paper, we work with the natural units, setting the speed of light $c$ and Planck constant $\hbar$ to unity.

\section{Setup}
\label{sec:basic_eq}

\subsection{Schr\"odinger-Poisson equation}
\label{subsec:S-P_eq}

Our starting point is to write down the basic equation for FDM in a cosmological background. Ignoring the self-interaction terms, the scalar field dark matter is generically described by the Klein-Gordon equation:
\begin{align}
 \Box\phi-\mass^2\phi=0
\label{eq:K-G_equation}
\end{align}
with $\mass$ being the mass of FDM. Consider the spacetime metric for the longitudinal gauge. Assuming the flat geometry, we have 
\begin{align}
 ds^2=-(1-2\Psi)dt^2+a^2(1-2\Psi)\,\delta_{ij}dx^idx^j.
\end{align}
Throughout the paper, we work with the weak-field limit, $|\Psi|\ll1$, and take the nonrelativistic limit of scalar field, introducing the following functional form:
\begin{align}
 \phi=\frac{1}{\sqrt{2\,\mass}\,a^{3/2}}\bigl[\varphi\,e^{-i\,\mass\,t}+\mbox{c.c}\bigr].
\end{align}
Since we are interested in the perturbations deep inside the Hubble horizon, we can make use of the approximation such that $H/\mass\sim\epsilon\sim k/\mass$ and $\Psi\sim\epsilon^2$ with $\epsilon$ being a small parameter, i.e., $\epsilon\ll1$. Then, the governing equation for the scalar field at Eq.~(\ref{eq:K-G_equation}) is shown to be reduced to the Schr\"odinger-Poisson system (e.g., Refs.~\cite{Marsh_review2016,Ferreira_review2020,Hui_review2021}). Adopting the natural units$(\hbar=c=1)$, we have
\begin{align}
 i\frac{\partial}{\partial t}\varphi=\Bigl[-\frac{1}{2\mass\,a^2}\nabla_x^2+\mass\,\Psi\Bigr]\varphi.
\label{eq:S-P_eq}
\end{align}
This has to be solved with the Poisson equation below:
\begin{align}
\frac{1}{a^2} \nabla^2_x\Psi=4\pi\,G\,\delta \rho_\phi,
\label{eq:Poisson_eq}
\end{align}
where the density fluctuation $\delta\rho_\phi$ is given by
\begin{align}
 \delta \rho_\phi \equiv \frac{\mass}{a^3}\big\{|\varphi|^2-\langle|\varphi|^2\rangle \}.
\label{eq:def_delta_rho_phi}
\end{align}
Equations~(\ref{eq:S-P_eq})$\sim$(\ref{eq:def_delta_rho_phi}) are the basic equations to describe the cosmic structure in FDM model. Note that taking the large $\mass$ limit, these equations have been also used to describe the structure formation in the CDM model \cite{Widrow_Kaiser1993}, in particular, as an alternative way to solve the Vlasov-Poisson equation (e.g., Refs.~\cite{Uhlemann_Kopp_Haugg2014,Ulhemann_Rampf_Gosenca_Hahn2019,Garny_Konstandin_Rubira2020}).

\subsection{FDM in a spherical halo}
\label{subsec:spherical}

A primary goal of this paper is to derive analytically the soliton core structure and to address its relation to the halo properties. Numerical simulations show that the soliton core forms at the center of halos, and the size and mass of the solitons are small enough compared to those of halos. Thus, at a first-order approximation for analytical study, we shall ignore the self-gravity of the soliton core, and treat the surrounding halo as a background. The impact of the soliton self-gravity will be later discussed (see Sec.~\ref{sec:discussion}). It is to be noted that halos found in simulations have large fluctuations in density, forming granular structures. While these fluctuations affect the soliton core and induce the motion and oscillations through the interaction with them, these nonstationary behaviors can be understood as the results of wave interference and are well described by a superposition of the ground (soliton) and excited states in a fixed potential \cite{Li_Hui_Yavetz2021, Luna_etal2022}. In this respect,constructing the eigenstates under a smooth halo background would be an important first step to describe the dynamical features of the soliton core.

\begin{figure*}[t]
 \includegraphics[width=10.5cm,angle=0]{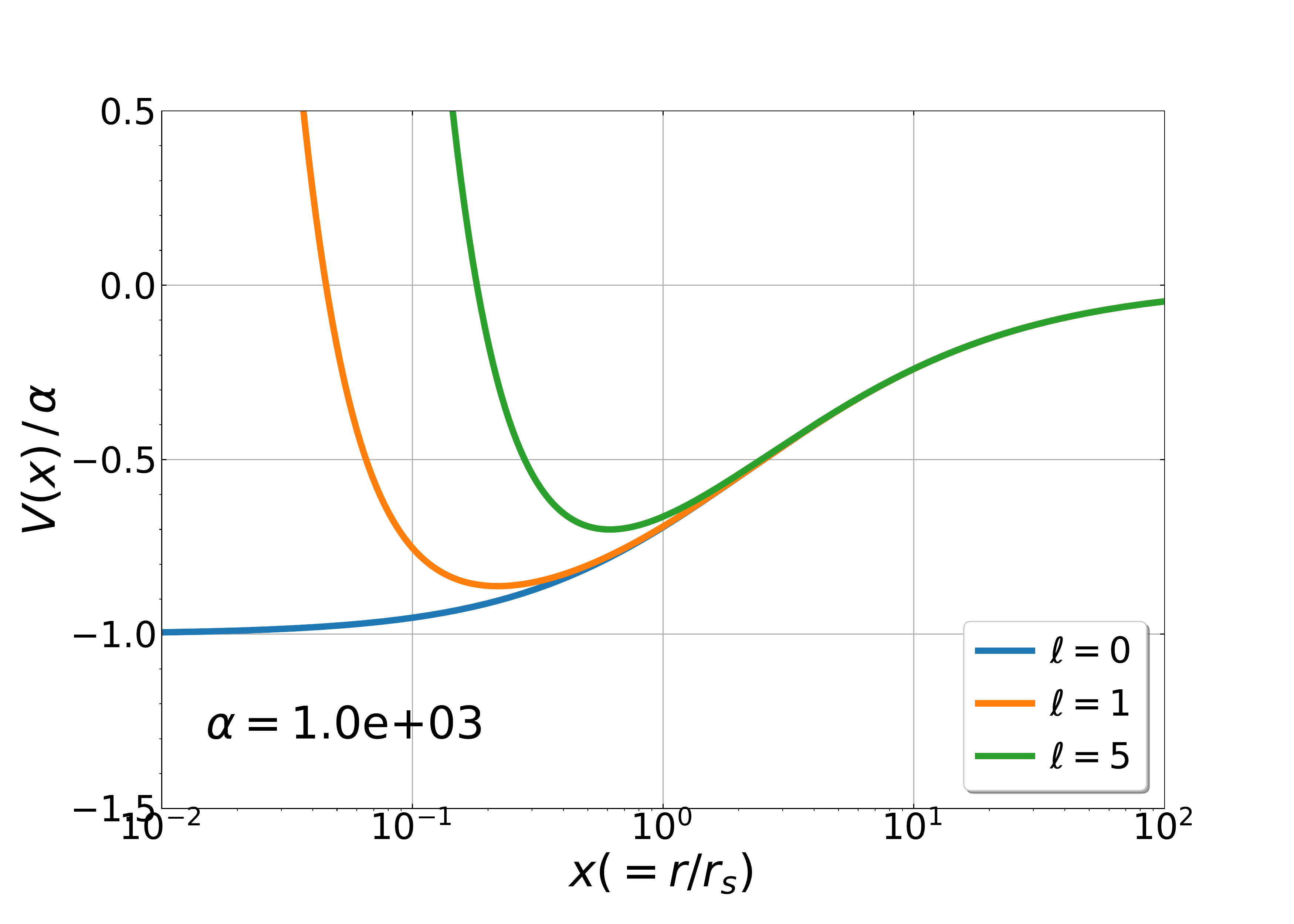}
 \includegraphics[width=6cm,angle=0]{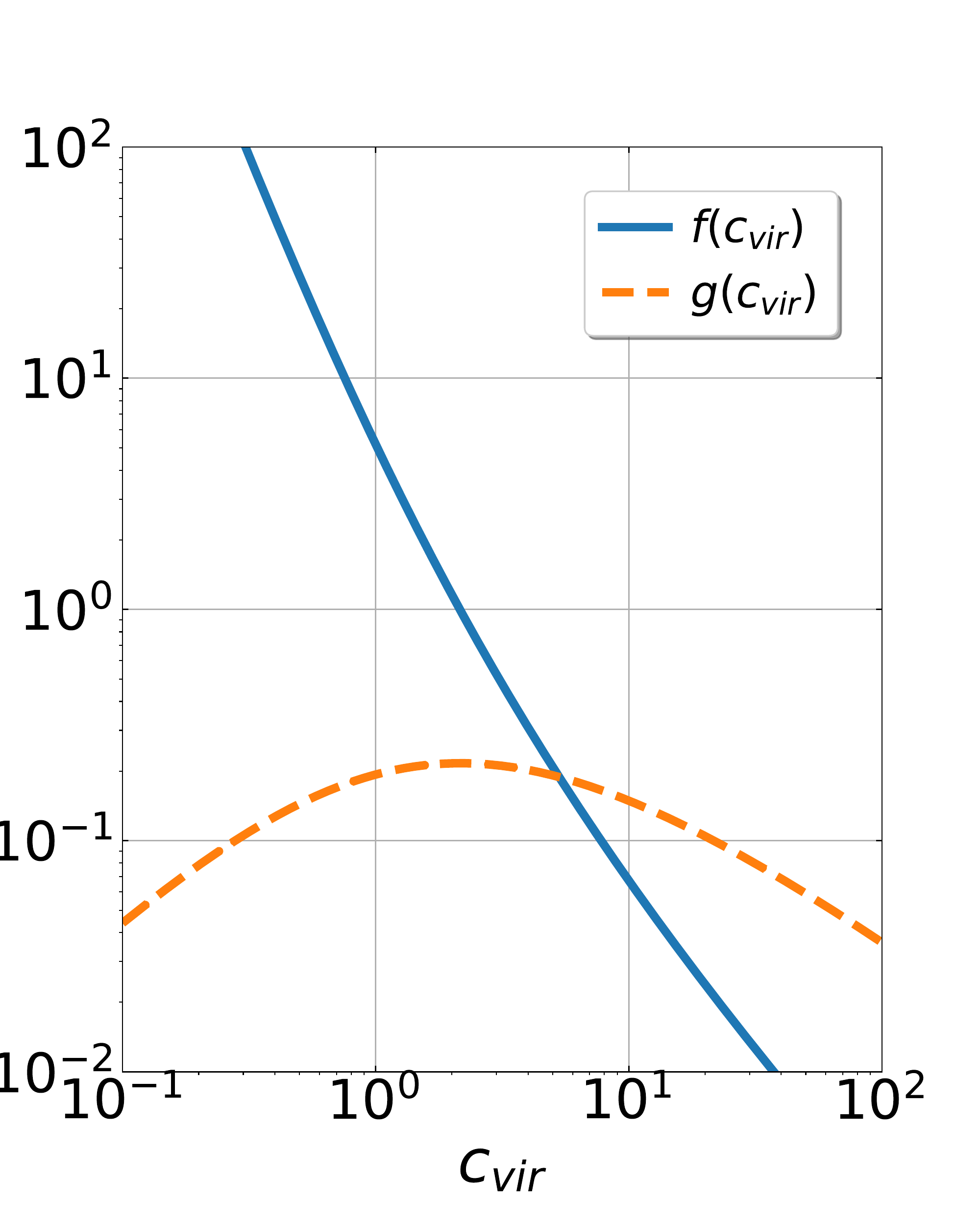}
\caption{{\it Left}: effective potential of the Schr\"odinger-Poisson equation in a NFW halo [see Eq.~(\ref{eq:stationary-SP_eq2})] normalized by $\alpha$, i.e., $V(x)/\alpha \equiv -\log(1+x)/x + \ell(\ell+1)/x^2/\alpha$. The results are plotted against the dimensionless radius, $x\equiv r/\rs$. Blue, orange, and green lines are the cases with $\ell=0$, $1$ and $5$, respectively. Here, we set the parameter $\alpha$ to $10^3$ for illustration.
 {\it Right}: functions $f(\cvir)$ and $g(\cvir)$, defined at Eqs.(\ref{eq:def_f_func}) and (\ref{eq:def_g_func}), as a function of concentration parameter $\cvir$.
\label{fig:potential_func_cvir}
}
\end{figure*}

Based on these considerations, we consider a spherically symmetric halo with the density profile described by the Navarro-Frenk-While (NFW) profile \cite{Navarro_Frenk_White1996,Navarro_Frenk_White1997}\footnote{We are interested in the region where the density is much larger than the mean density, and the contribution of the term $\langle|\varphi|^2\rangle$ in Eq.~(\ref{eq:def_delta_rho_phi}) is ignored. }:
\begin{align}
 \mass\,|\varphi|^2&\simeq\rho_{\rm NFW}(r)
\equiv\frac{\rhos}{(r/\rs)(1+r/\rs)^2}.
\label{eq:rho_NFW}
\end{align}
Note that the density $\rho_{\rm NFW}$ defined above differs from the one frequently used in the literature, and a factor of $1/a^3$ is factored out. Plugging Eq.~(\ref{eq:rho_NFW}) into Eq.~(\ref{eq:def_delta_rho_phi}), the Poisson equation at Eq.~(\ref{eq:Poisson_eq}) is analytically solved, and the potential is expressed as (e.g., see Appendix D of Ref.~\cite{Saga_etal2020})\footnote{Equation~(\ref{eq:NFW_potential}) is obtained under the assumption that the density fluctuation $\delta \rho_\phi$ follows the NFW profile even outside the virial radius. While this is not in reality true in both the observations and simulations, the modification of the outer profile hardly changes the inner potential structure, and hence the analysis based on Eq.~(\ref{eq:NFW_potential}) is relevant to describe the soliton core structure.}
\begin{align}
 \Psi=-4\pi\,G\,\frac{\rhos\,\rs^2}{a}\,\frac{\log(1+r/\rs)}{r/\rs}.
\label{eq:NFW_potential}
\end{align}
Treating this halo potential as a background, solving the Schr\"odinger-Poisson (S-P) equation at Eq.~(\ref{eq:Poisson_eq}) is reduced to a linear eigenvalue problem. Writing the wave function in the form as
\begin{align}
\varphi =\sum_{n \ell m}  u_{n\ell}(r)\,Y_{\ell m}(\theta,\phi)\,e^{-i\,\epsilon \tau},
\end{align}
where the quantity $\tau$ is the new time variable defined by $\tau\equiv\int^t dt'/\{a(t')\}^2$, Eq.~(\ref{eq:S-P_eq}) with the potential (\ref{eq:NFW_potential}) yields
\begin{widetext}
\begin{align}
-\frac{1}{2\mass} \frac{1}{r^2}\frac{d}{dr}\Bigl(r^2\frac{du_{n\ell}(r)}{dr}\Bigr)+\Bigl[-4\pi\,G\,\mass\,\rhos\,\rs^2\,a\,\frac{\log(1+r/\rs)}{r/\rs}
+\frac{\ell(\ell+1)}{2\mass\,r^2}
\Bigr] u_{n\ell}(r)=\epsilon\,u_{n\ell}(r).
\label{eq:stationary-SP_eq}
\end{align}
Equation~(\ref{eq:stationary-SP_eq}) is further rewritten in a simplified form by 
introducing dimensionless quantities: 
\begin{align}
 x\equiv \frac{r}{\rs},\qquad 
\mathcal{E}\equiv 2\mass\,\rs^2\epsilon,\qquad
\alpha \equiv 8\pi\,G\,\mass^2\,\rhos\,\rs^4\,a.
\label{eq:parameters}
\end{align}
We have
\begin{align}
& -\frac{1}{x^2}\frac{d}{dx}\Bigl(x^2\frac{du_{n\ell}(x)}{dx}\Bigr)+\Bigl[-\alpha\,\frac{\log(1+x)}{x}+\frac{\ell(\ell+1)}{x^2}\Bigr]u_{n\ell}(x)
=\mathcal{E}\,u_{n\ell}(x),
\label{eq:stationary-SP_eq2}
\end{align}
Here, the quantity $\alpha$ characterizes the depth of the potential well, and it typically takes a large value. Using the quantities characterizing the properties of halos in Appendix \ref{app: NFW_profile}, we obtain
\begin{align}
\alpha &=7.21\times 10^2\,
\Bigl(\frac{\mass}{10^{-22}{\rm eV}}\Bigr)^2\,
\Bigl(\frac{M_{\rm h}}{10^9\,M_\odot}\Bigr)^{4/3}
\Bigl(\frac{\Delta_{\rm vir}(z)}{200}\Bigr)^{-1/3}
\Bigl(\frac{\Omega_{\rm m,0}h^2}{0.147}\Bigr)^{-1/3}\,\frac{f(\cvir)}{1+z}
\label{eq:alpha_value}
\end{align}
\end{widetext}
with the function $f$ defined by
\begin{align}
f(\cvir)\equiv \frac{1}{\cvir\{\ln(1+\cvir)-\cvir/(1+\cvir)\}}.
\label{eq:def_f_func}
\end{align}
The quantity $\cvir$ is the concentration parameter, $\cvir\equiv r_{\rm vir}/\rs$, with $r_{\rm vir}$ being the virial radius of the halo having the mass of $M_{\rm h}$.  For typical values of $\cvir\sim2-10$, the function $f$ varies from $f\sim1$ to $0.07$ (see right panel of Fig.~\ref{fig:potential_func_cvir}). Left panel of Fig.~\ref{fig:potential_func_cvir} shows the effective potential normalized by the parameter $\alpha$, i.e., $V(x)/\alpha\equiv-\log(1+x)/x+\ell(\ell+1)/x^2/\alpha$, where the parameter $\alpha$ is set to $10^3$. Note that the parameter $\alpha$ is not strictly a constant value, but it depends explicitly on the redshift. In this respect, Eqs.~(\ref{eq:stationary-SP_eq}) or (\ref{eq:stationary-SP_eq2}) cannot be reduced to a stationary problem. Nevertheless, estimating a typical timescale for the wave function, defined by $\tau_{\rm osc}\equiv 2\pi/|\epsilon|$, gives 
\begin{widetext}
\begin{align}
\tau &= \frac{1}{4\pi\,G\,\mass\,\rhos\,\rs^2}\,\Bigl(\frac{|\mathcal{E}|}{\alpha}\Bigr)^{-1}
\nonumber
\\
&=1.34\times10^7\,\,[{\rm year}]\,\,\Bigl(\frac{|\mathcal{E}|}{\alpha}\Bigr)^{-1}
\Bigl(\frac{\mass}{10^{-22}\,{\rm eV}}\Bigr)^{-1}\,
\Bigl(\frac{M_{\rm h}}{10^9\,M_\odot}\Bigr)^{-2/3}
\Bigl(\frac{\Delta_{\rm vir}}{200}\Bigr)^{-1/3}
\Bigl(\frac{\Omega_{\rm m,0}h^2}{0.147}\Bigr)^{-1/3}\,\Bigl\{\frac{g(\cvir)}{0.1}\Bigr\}
\label{eq:def_tau_osc}
\end{align}
\end{widetext}
with the function $g(\cvir)$ defined by
\begin{align}
 g(\cvir)\equiv\frac{1}{\cvir^2\,f(\cvir)}, 
\label{eq:def_g_func}
\end{align}
which typically has $g\sim0.15-0.2$ for a broad range of $\cvir$ (see right panel of Fig.~\ref{fig:potential_func_cvir}). As we will see later, the typical value of the energy level is given by $|\mathcal{E}|\sim\alpha$ (Sec.~\ref{subsec:comparison}). Hence, the oscillation period of the wave function is sufficiently shorter than the cosmological timescales, and thus we can ignore the time dependence of the parameter $\alpha$.

\section{Analytical solutions for $\ell=0$}
\label{sec:Analytic_solutions_ell0}

In this section, in order to describe the soliton core structure and its properties, we focus on the cases with zero angular momentum ($\ell=0$), and obtain an approximate solution for the radial part of the S-P equation. In Appendix \ref{sec:Analytic_solutions_non_zero_ell},  we also construct the approximate eigenstates for the nonzero angular momentum, and the results are compared with numerical solutions.

\subsection{Constructing analytical eigenfunctions and eigenvalues}
\label{subsec:Langer_approx}

Hereafter, dropping the subscript $\ell$, we denote the radial wave function for $\ell=0$ by $u_n$. Introducing the new function $\tilde{u}_n\equiv x\,u_n$ and setting the angular momentum $\ell$ to zero, we rewrite Eq.~(\ref{eq:stationary-SP_eq2}) with the normal form below:
\begin{align}
 \frac{d^2\tilde{u}_n(x)}{dx^2}+\alpha\,g_0(x)\tilde{u}_n(x)=0\,;\quad
g_0(x)\equiv \frac{\log(1+x)}{x}+\frac{\mathcal{E}}{\alpha}.
\label{eq:SP_eq3_ell0}
\end{align}
We shall construct the bound-state solutions satisfying the boundary condition\footnote{The inner boundary condition $\tilde{u}(0)=0$ ensures that the mass profile, $M(x)\propto \int_0^x |u(x')|^2\,x'^2dx'=\int_0^x |\tilde{u}(x')|^2dx'$, is regular at $x=0$, and we have $M(0)=0$.}: 
\begin{align}
 \tilde{u}_n(0)=0,\qquad \tilde{u}_n(\infty)=0.
\label{eq:boundary_condition}
\end{align}

To obtain an approximate solution for the bound-state FDM distribution, we first notice that the function $g_0(x)$ in Eq.~(\ref{eq:SP_eq3_ell0}) is a monotonically decreasing function of $x$, and we have $g_0(0)=1+\mathcal{E}/\alpha$ and $g_0(\infty)=\mathcal{E}/\alpha$. Since the eigenvalue $\mathcal{E}$ for the bound-state solution must be negative ($\mathcal{E}<0$), the function $g_0$ should have a single zero-crossing (or turning) point. Denoting its position by $x_c$, we have
\begin{align}
 g_0(x_c)=\frac{\log(1+x_c)}{x_c}+\frac{\mathcal{E}}{\alpha}=0.
\label{eq:turning_point}
\end{align}
Note that the above equation implies that the position $x_c$ depends on the eigenvalue $\mathcal{E}$. The eigenvalues are given as a set of discrete values associated to the eigenfunction $\tilde{u}$, and will be later determined by the boundary condition at Eq.~(\ref{eq:boundary_condition}).

Keeping in mind the fact that Eq.~(\ref{eq:SP_eq3_ell0}) has a single turning point, we consider the following transformation, known as the Liouville-Green transformation (see e.g., \cite{Nayfeh_1981}):
\begin{align}
 z=p(x),\qquad v_n(z)=\sqrt{p'(x)}\,\tilde{u}_n(x).
\label{eq:Liouville-Green_transformation}
\end{align}
The prime denotes the derivative with respect to $x$. With this transformation, Eq.~(\ref{eq:SP_eq3_ell0}) becomes
\begin{align}
 \frac{d^2v_n(z)}{dz^2}+\Bigl[\alpha\frac{g_0(x)}{\{p'(x)\}^2}+\delta\Bigr]v_n(z)=0,
\label{eq:SP_eq4_ell0}
\end{align}
where the quantity $\delta$ is defined by
\begin{align}
\delta\equiv\frac{q}{\{p'\}^2}\Bigl(\frac{1}{q}\Bigr)'',\,\, \quad q=\sqrt{p'}.
\label{eq:def_delta}
\end{align}
Here, we choose the function $p$ so as to satisfy\footnote{This choice is known as the Langer's transformation \cite{Langer1932}. See Chapter 14.6 of Ref.~\cite{Nayfeh_1981}}
\begin{align}
 \alpha\, \frac{g_0(x)}{\{p'(x)\}^2}=-z.
\label{eq:choice_p}
\end{align}
Recall from Eq.~(\ref{eq:Liouville-Green_transformation}) that $z$ is equal to $p$, the above equation is solved to give an explicit form of $p$ as follows. First we take the square root of both sides and obtain
\begin{align}
 p^{1/2}p'=\pm\alpha^{1/2}\,\sqrt{-g_0(x)}
\end{align}
for $x>x_c$. Upon separation of variables, this becomes
\begin{align}
 p^{1/2}dp=\pm\alpha^{1/2}\,\sqrt{-g_0(x)}\,dx.
\end{align}
Integrating once, we have
\begin{align}
  \frac{2}{3} z^{3/2}=\frac{2}{3} p^{3/2}=\pm \alpha^{1/2}\,\int_{x_c}^x\,\sqrt{-g_0(x')}\,dx'.
\label{eq:z_integral}
\end{align}
Here we set the lower value of the integral to $x_c$. Taking the positive sign of Eq.~(\ref{eq:z_integral}), we obtain
\begin{align}
 z=p(x)=\alpha^{1/3}\,\Bigl[\frac{3}{2}\int_{x_c}^x \sqrt{-g_0(x')}\,dx' \Bigr]^{2/3},
\label{eq:p_z_func_x_plus}
\end{align}
which is valid at $x>x_c$ (i.e., $z>0$). Similarly, we obtain the relation valid at $x\leq x_c$ (or $z\leq0$):
\begin{align}
 z=p(x)=-\alpha^{1/3}\,\Bigl[\frac{3}{2}\int^{x_c}_x \sqrt{g_0(x')}\,dx' \Bigr]^{2/3}.
\label{eq:p_z_func_x_minus}
\end{align}
With Eqs.~(\ref{eq:p_z_func_x_plus}) and (\ref{eq:p_z_func_x_minus}), the function $z$ yields a smooth and monotonically increasing function of $x$, and its sign changes from negative to positive at $x_c$. 

Using Eq.~(\ref{eq:choice_p}) and the relations given above, the differential equation given at Eq.~(\ref{eq:SP_eq4_ell0}) is further recast as
\begin{align}
\frac{d^2v_n(z)}{dz^2}-z\,v_n(z)=-\delta\,v_n(z). 
\end{align}
Equations~(\ref{eq:p_z_func_x_plus}) and (\ref{eq:p_z_func_x_minus}) ensure that the quantity $\delta$ defined at Eq.~(\ref{eq:def_delta}) is shown to be of the order of $\mathcal{O}(\alpha^{-2/3})$ for all values of $x$. Thus, in the limit of $\alpha\gg1$, we can safely ignore the right-hand side involving the factor $\delta$, and the above equation is reduced to 
\begin{align}
\frac{d^2v_n(z)}{dz^2}-z\,v_n(z)\simeq0.
\end{align}
The general solution of this equation is known to be expressed in terms of the Airy functions:
\begin{align}
 v_n(z)=\tilde{c}_1\,\mbox{Ai}(z) +\,\tilde{c}_2\,\mbox{Bi}(z)
\end{align}
with $\tilde{c}_1$ and $\tilde{c}_2$ being the integration constants. 
Note that the asymptotic behaviors of the Airy functions are (e.g., \cite{Abramowitz_Stegun1965})
\begin{widetext}
\begin{align}
 \mbox{Ai}(z)\sim\left\{
\begin{array}{ll}
{\displaystyle \frac{e^{-(2/3)z^{3/2}}}{2\sqrt{\pi}\,z^{1/4}} } & (z\to\infty )
\\
\\
{\displaystyle \frac{\sin\Bigl[(2/3)(-z)^{3/2}+\pi/4\Bigr]}{\sqrt{\pi}\,(-z)^{1/4}} }& (z\to-\infty )
\end{array}
\right.,
\quad
 \mbox{Bi}(z)\sim\left\{
\begin{array}{ll}
{\displaystyle \frac{e^{(2/3)z^{3/2}}}{\sqrt{\pi}\,z^{1/4}} } & (z\to\infty )
\\
\\
{\displaystyle \frac{\cos\Bigl[(2/3)(-z)^{3/2}+\pi/4\Bigr]}{\sqrt{\pi}\,(-z)^{1/4}} }& (z\to-\infty )
\end{array}
\right.
.
\label{eq:asymptotic_Airy_func}
\end{align}
\end{widetext}
Thus, imposing the boundary condition $\tilde{u}_n(\infty)=0$, the constant $\tilde{c}_2$ must be zero. Another boundary condition, $\tilde{u}_n(0)=0$, leads to the following condition,
\begin{align}
 \mbox{Ai}\bigl(z(0)\bigr)=0, 
\label{eq:eigenvalue_Airy}
\end{align}
where we used the fact that both of the functions $z(x)$ and $g_0(x)$ are nonvanishing at $x=0$.

Equation~(\ref{eq:eigenvalue_Airy}) determines the eigenvalues $\mathcal{E}$ through the function $z(0)$ given at Eq.~(\ref{eq:p_z_func_x_minus}), in which the integrand $\sqrt{g_0(x)}$ manifestly depends on $\mathcal{E}$ [see Eq.~(\ref{eq:SP_eq3_ell0}] for the definition of $g_0(x)$). It is the transcendental equation and has to be evaluated numerically to get the eigenvalues. Although this can be easily and quickly done with the standard numerical techniques, we can derive an analytical expression for energy eigenvalues by further employing an approximation. Since the quantity $z$ is of the order of $\mathcal{O}(\alpha^{1/3})$ and $z<0$ at $x=0$, we can make use of the asymptotic form of the Airy function $\mbox{Ai}$ by taking the limit of $\alpha\gg1$. Then, Eq.~(\ref{eq:eigenvalue_Airy}) is rewritten with [see Eq.~(\ref{eq:asymptotic_Airy_func})]
\begin{align}
 \frac{1}{\sqrt{\pi}\,\{-z(0)\}^{1/4}}\sin\Bigl(\frac{2}{3}\{-z(0)\}^{3/2}+\frac{\pi}{4}\Bigr)=0.
\end{align}
Note that Eq.~(\ref{eq:p_z_func_x_minus}) implies $z(0)\ne0$. Thus, the condition given above is recast as [using Eq.~(\ref{eq:p_z_func_x_minus})]
\begin{align}
& \frac{2}{3}\{-z(0)\}^{2/3}+\frac{\pi}{4}=n\,\pi\,\,
\nonumber
\\
&\qquad\Longleftrightarrow 
\,\,\alpha^{1/2}\int_0^{x_c}\sqrt{g_0(x')}\,dx'=\Bigl(n-\frac{1}{4}\Bigr)\pi
\label{eq:eigenvalue_semi-analytic}
\end{align}
with the variable $n$ being the positive integer (i.e., $n=1,\,2,\,\cdots$). Assuming further $x_c\ll1$ and expanding the function $g_0(x)$ as $g_0(x)\simeq 1-x/2+\mathcal{E}/\alpha$, the turning point is easily evaluated to give $x_{c}=2(1+\mathcal{E}/\alpha)$. Substituting the relations into the above, the integration is analytically performed, and  Eq.~(\ref{eq:eigenvalue_semi-analytic}) becomes
\begin{align}
\alpha^{1/2}\, \frac{4}{3}\Bigl(1+\frac{\mathcal{E}}{\alpha}\Bigr)^{3/2}=\Bigl(n-\frac{1}{4}\Bigr)\pi.
\end{align}
Solving the above equation with respect to $\mathcal{E}$, we finally get an analytical expression for the energy eigenvalues: 
\begin{align}
 \Bigl(\frac{\mathcal{E}}{\alpha}\Bigr)_{\rm approx}=-1+\alpha^{-1/3}\Bigl\{\frac{3}{4}\Bigl(n-\frac{1}{4}\Bigr)\pi\Bigr\}^{2/3}.
\label{eq:eigenvalue_approximation}
\end{align}
Note that the $n=1$ implies the ground state. The above analytical expression is in general valid as long as we consider $\alpha \gg1$ and for a small integer $n$.

In Table \ref{tab:eigenvalues}, normalizing the energy eigenvalues by $\alpha$, 
we list their numerical values for the lowest five eigenstates. Here, we specifically show the results of $\alpha=10^3$. Table \ref{tab:eigenvalues} demonstrates how well the analytical estimation at Eq.~(\ref{eq:eigenvalue_approximation}) can work. Compared to the results obtained from numerical calculation and Eq.~(\ref{eq:eigenvalue_Airy}), the eigenvalues from (\ref{eq:eigenvalue_approximation}) start to deviate as increasing $n$. On the other hand, results obtained by solving the transcendental equation (\ref{eq:eigenvalue_Airy}) are surprisingly accurate even for $n=5$. We will see in next subsection how this semianalytical estimation works well in more general cases with various values of $\alpha$. 
\begin{table}[t]
 \caption{Comparison of analytically and numerically estimated eigenvalues for the zero angular-momentum case. Setting the parameter $\alpha$ to $10^3$, numerical values of $\mathcal{E}$ are summarized by normalizing them by $\alpha$, particularly for the lowest five eigenstates.
\label{tab:eigenvalues}}
\begin{ruledtabular}
\begin{tabular}{lccc}
  $\mathcal{E}/\alpha$ & Numerical  & Eq.~(\ref{eq:eigenvalue_Airy})  & Eq.~(\ref{eq:eigenvalue_approximation}) 
\\
\hline
$n=1$  & -0.86680 & -0.86687 & -0.85383
\\
$n=2$  & -0.78318 & -0.78323 & -0.74286
\\
$n=3$  & -0.72307 & -0.72311 & -0.65244
\\
$n=4$  & -0.67531 & -0.67535 & -0.57261
\\
$n=5$  & -0.63553 & -0.63557 & -0.49965
\\

\hline
\end{tabular}
\end{ruledtabular}
\end{table}

To sum up, the analytical solution for the bound-state radial wave function which satisfies the boundary conditions $\tilde{u}(0)=0=\tilde{u}(\infty)$ is given by 
\begin{widetext}
\begin{align}
 \tilde{u}_{n}(x)=x\,u_n(x)=\left\{
\begin{array}{ll}
{\displaystyle \frac{\{z(x)\}^{1/4}}{\{-g_0(x)\}^{1/4}}\,\mbox{Ai}(z(x))}\,;\quad 
z(x)=\alpha^{1/3}\Bigl[\frac{3}{2}\int_{x_c}^x\sqrt{-g_0(x')}\,dx'\Bigr]^{2/3},
 & (x>x_c)
\\
\\
{\displaystyle \frac{\{-z(x)\}^{1/4}}{\{g_0(x)\}^{1/4}}\,\mbox{Ai}(z(x))}\,;\quad 
z(x)=-\alpha^{1/3}\Bigl[\frac{3}{2}\int_x^{x_c}\sqrt{g_0(x')}\,dx'\Bigr]^{2/3},
 & (0\leq x\leq x_c)
\end{array}
\right.
\label{eq:eigen_func_analytical_ell0}
\end{align}
\end{widetext}
with the function $g_0$ defined at Eq.~(\ref{eq:SP_eq3_ell0}). The eigenvalue $\mathcal{E}$ is obtained by solving Eq.~(\ref{eq:eigenvalue_Airy}) for a given $\alpha$, and the turning point $x_c$ is determined by $g_0(x_c)=0$. Note that the analytical expressions given above is not normalized, and for a proper definition, it has to be divided by the normalization constant $\mathcal{N}$, defined by $\mathcal{N}\equiv \{\int dx' |\tilde{u}_n(x')|^2\}^{1/2}$.

\begin{figure*}[tb]
 \includegraphics[width=18cm,angle=0]{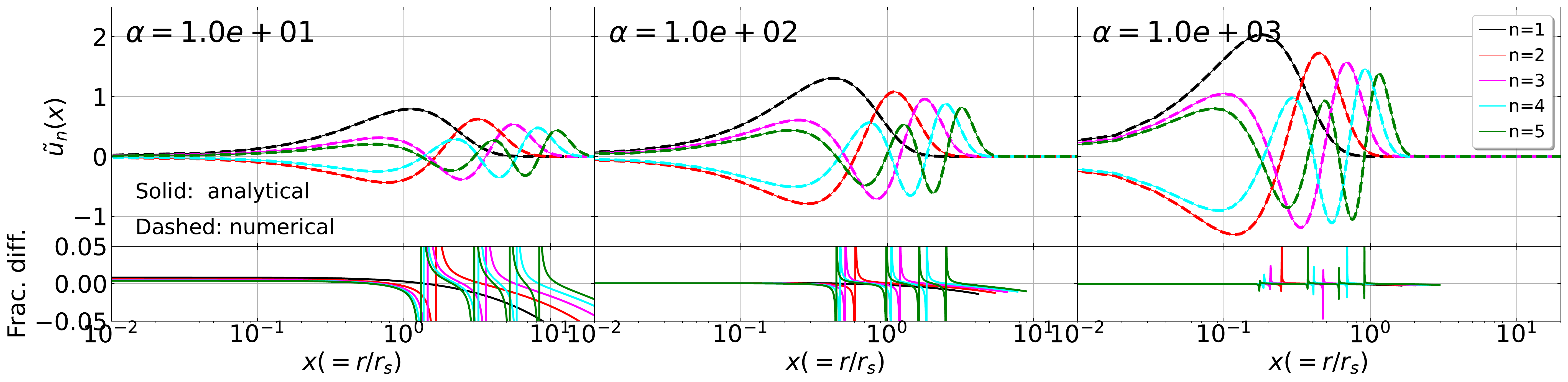}
\caption{Eigenfunctions for radial wave function $\tilde{u}_{n}(x)$ for $\alpha=10$ (left), $10^2$ (middle), and $10^3$ (right). Results of the lowest five eigenstates are plotted as a function of dimensionless radius $x=r/\rs$. shown. In the upper panels, while the thin solid lines are the analytical results based on Eq.~(\ref{eq:eigen_func_analytical_ell0}), thick dashed lines represent the numerical solutions obtained by solving the matrix eigenvalue problem. Note that all the results are normalized. On the other hand, the bottom panels show the fractional difference between the analytical and numerical results defined by $\Delta \tilde{u}_n/\tilde{u}_{n,{\rm numerical}}$ with $\Delta\tilde{u}_n=\tilde{u}_{n,{\rm analytical}}-\tilde{u}_{n,{\rm numerical}}$, where the functions $\tilde{u}_{n,{\rm analytical}}$ and $\tilde{u}_{n,{\rm numerical}}$ denote the analytical and numerical eigenstates, respectively. Here, to avoid the division by zero, we stop plotting the results when the amplitude of the wave functions becomes smaller than $10^{-9}$ at the outer part. 
\label{fig:eigen_func_alpha10_100_1000} }

\vspace*{0.5cm}

\includegraphics[width=18cm,angle=0]{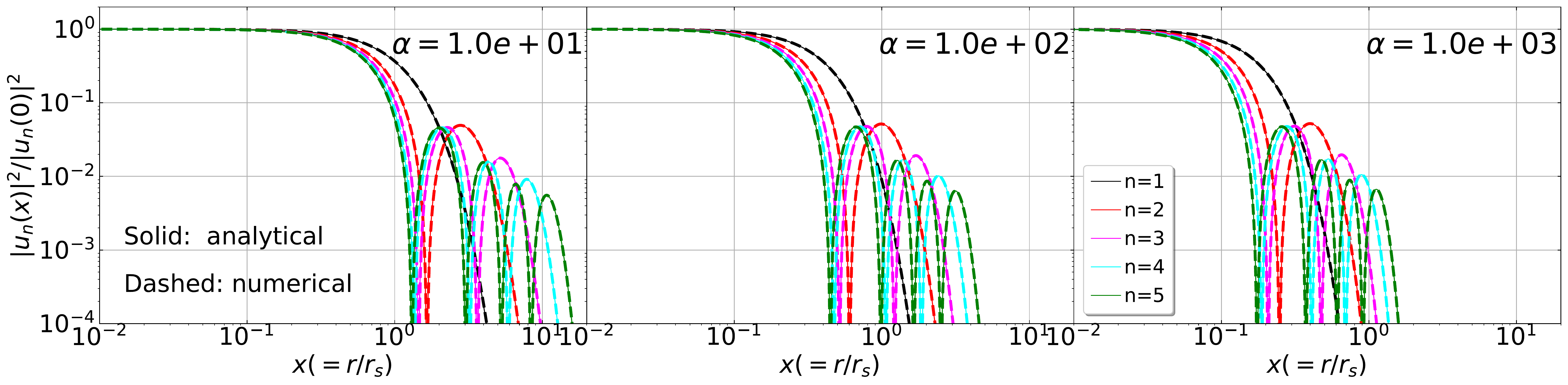}
\caption{Same as Fig.~\ref{fig:eigen_func_alpha10_100_1000}, but we here plot the square of the wave function $u_n(x)=x\,\tilde{u}_n(x)$, normalized by the one evaluated at the origin, i.e., $|u_n(x)|^2/|u_n(0)|^2$.  Note that the plotted results correspond to the density profile normalized by the central density, $\rho(x)/\rho(0)$. \label{fig:eigen_func_log_alpha10_100_1000} }
\end{figure*}

\subsection{Comparison with numerical solutions}
\label{subsec:comparison}

Having obtained the analytical expressions for eigenvalues and eigenfunctions,  
we now compare them with numerical solutions. Based on the standard technique to solve the stationary problem of Schr\"odinger equation, we discretize Eq.~(\ref{eq:SP_eq3_ell0}). Then, solving differential equation under the boundary conditions at Eq.~(\ref{eq:boundary_condition}) is reduced to a matrix eigenvalue problem in the following form (e.g., Ref.~\cite{Luna_Zagorac_etal2021}): 
\begin{widetext}
\begin{align}
\left(
\begin{array}{ccccc}
2/\Delta^2 + V(x_1)& -1/\Delta^2 & \cdots &0 &0   \\
-1/\Delta^2& 2/\Delta^2 + V(x_2) & -1/\Delta^2 &\cdots &0  \\
\vdots & \ddots & \ddots &\ddots & \vdots  \\
0 & \ddots & -1/\Delta^2 & 2/\Delta^2 + V(x_{n-1})  & -1/\Delta^2 \\
0 & \cdots & 0 & -1/\Delta^2 & 2/\Delta^2 + V(x_n)  \\
\end{array}
\right)\left(
\begin{array}{c}
\tilde{u}(x_1) \\
\tilde{u}(x_2) \\
    \vdots        \\
\tilde{u}(x_{n-1}) \\
\tilde{u}(x_n) 
\end{array}
\right) = \mathcal{E}\left(
\begin{array}{c}
\tilde{u}(x_1) \\
\tilde{u}(x_2) \\
    \vdots        \\
\tilde{u}(x_{n-1}) \\
\tilde{u}(x_n) 
\end{array}
\right)
\label{eq:matrix_eigenvalue_problem}
\end{align}
\end{widetext}
with the quantity $\Delta$ and the function $V(x)$, respectively, defined by $\Delta\equiv x_{i+1}-x_i$ and $V(x)=-\alpha\,\log(1+x)/x$. 

Figures~\ref{fig:eigen_func_alpha10_100_1000} and \ref{fig:eigen_func_log_alpha10_100_1000} show the wave functions of the lowest five eigenstates (i.e., $n=1,\cdots,5$) for the parameters $\alpha=10$ (left), $10^2$ (middle), and $10^3$ (right). 
In upper panels of Fig.~\ref{fig:eigen_func_alpha10_100_1000}, thick dashed lines are the numerical results of the function $\tilde{u}_n$, which are obtained by setting the inner and outer boundaries to $x_1=0$ and $x_n=50$ for $\alpha=10$ and $20$ for $\alpha=10^2$ and $10^3$ with the number of grids $n=10^4$. These results are compared to the analytical results depicted as thin solid lines, with the amplitude of each eigenfunction properly normalized. Bottom panels of Fig.~\ref{fig:eigen_func_alpha10_100_1000} plot the fractional difference between the analytical and numerical results, defined by $(\tilde{u}_{n,{\rm analytical}} - \tilde{u}_{n,{\rm numerical}})/\tilde{u}_{n,{\rm numerical}}$. On the other hand, Fig.~\ref{fig:eigen_func_log_alpha10_100_1000} shows the square of the wave function $u_n=x\,\tilde{u}_n$, normalized it by the one evaluated at the origin, i.e., $|u_n(x)|^2/|u_n(0)|^2$, which corresponds to the density profile normalized by the central density, $\rho(x)/\rho(0)$. The meaning of line types and colors are the same as those shown in the upper panels of Fig.~\ref{fig:eigen_func_alpha10_100_1000}.

Clearly the agreement between analytical numerical results is 
excellent even for a small value of $\alpha$, with which we naively expect the analytical prediction to be inaccurate. A closer look at the fractional difference reveals a discrepancy at larger radii $x$ for small $\alpha$, where spiky features arising from the zero-crossing points also become prominent. Nevertheless, these behaviors appear manifest only when the wave functions fall off and approach zero, thus giving no serious impact even from the quantitative point of view. Indeed, looking at the density profiles plotted in logarithmic scales (Fig.~\ref{fig:eigen_func_log_alpha10_100_1000}), we hardly see a difference between analytical and numerical results. 

\begin{figure*}[tb]
 \includegraphics[width=8.5cm,angle=0]{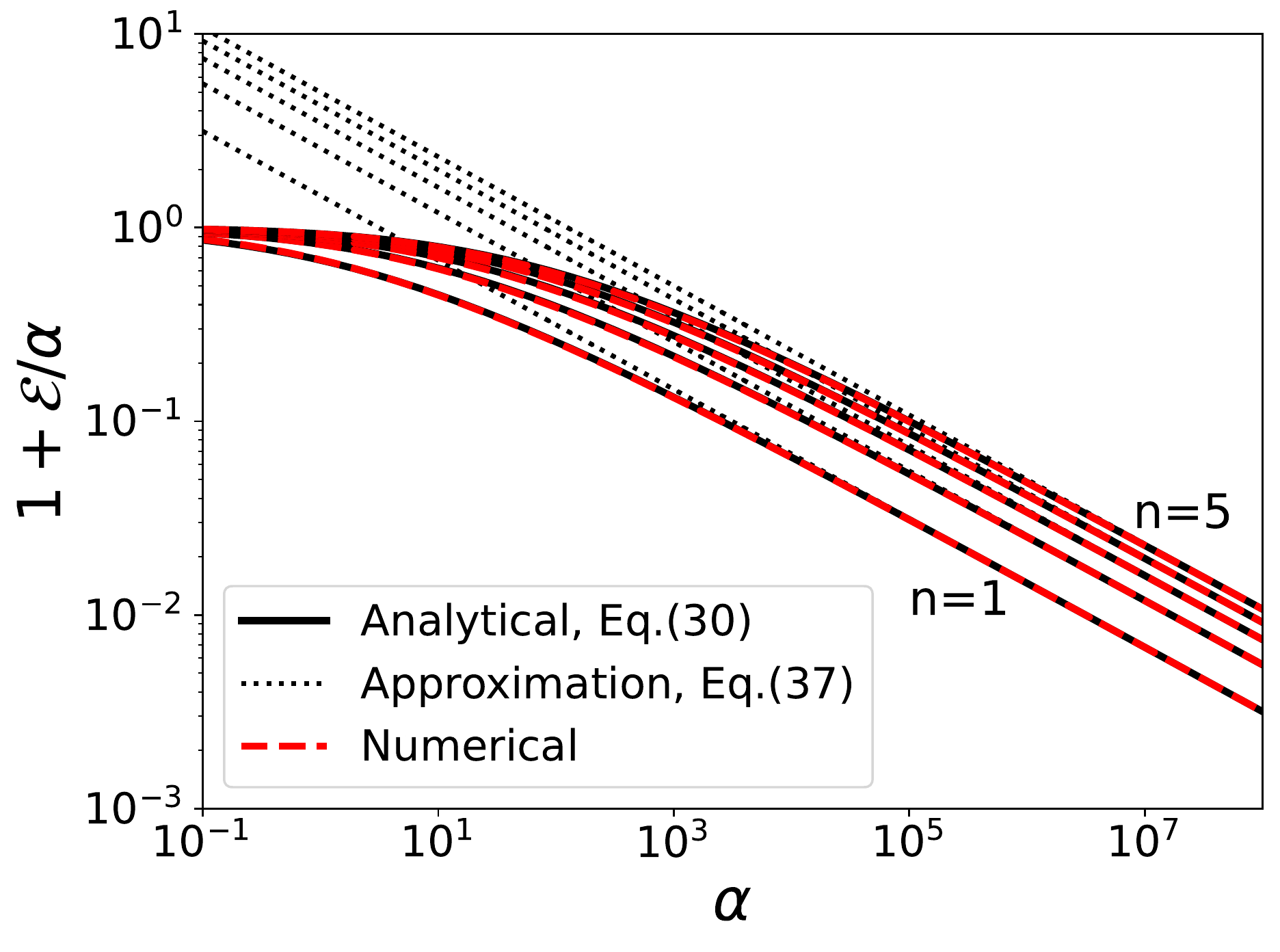}
 \includegraphics[width=8.5cm,angle=0]{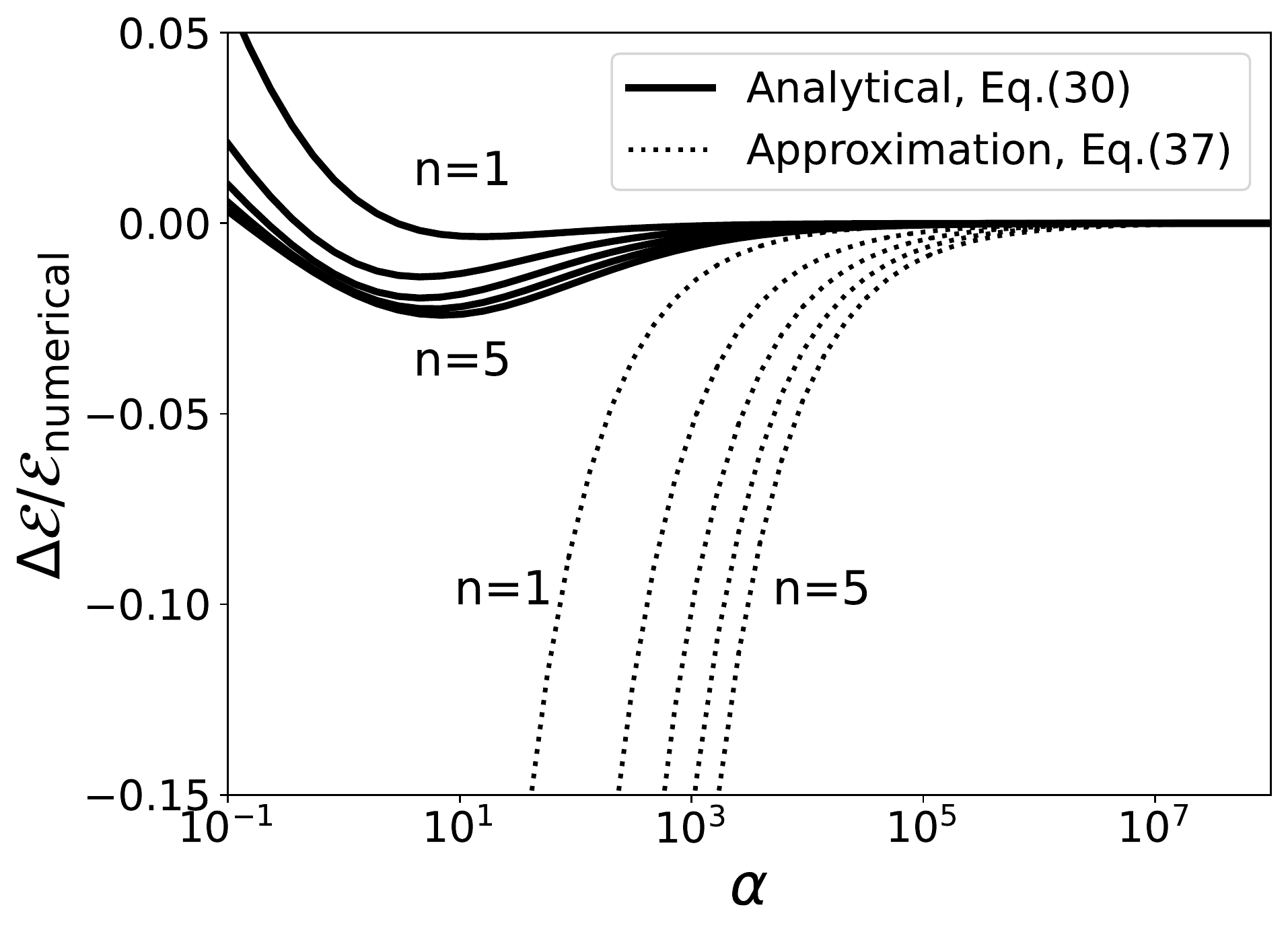}
\hspace*{0.5cm}
\caption{{\it Left}: energy eigenvalue plotted as a function of $\alpha$. To be precise, the plotted results are $1+\mathcal{E}/\alpha$, and results for the lowest five eigenstates are shown. Lines from bottom to top indicate the results from $n=1$ to $5$. {\it Right}: fractional error of the analytically estimated values of the eigenvalue $\mathcal{E}$,  $\Delta\mathcal{E}/\mathcal{E}_{\rm numerical}$. Again, results for the lowest five eigenstates are shown. 
\label{fig:eigenvalues_n1-5}
}
\end{figure*}

Next look at the eigenvalues. Table \ref{tab:eigenvalues} summarizes the results in the case of $\alpha=10^3$. Further, we plot in Fig.~\ref{fig:eigenvalues_n1-5} the eigenvalues for the lowest five eigenstates as a function of $\alpha$. Here, the eigenvalues computed from Eqs.~(\ref{eq:eigenvalue_Airy}) and (\ref{eq:eigenvalue_approximation}) are shown, depicted as solid and dotted lines, respectively. Again, we see an excellent agreement between numerical and analytical estimations. It is surprising that the analytical estimation with Eq.~(\ref{eq:eigenvalue_Airy}) remains accurate even at $\alpha\lesssim1$, and reproduce numerical results. Thus, we conclude that the solutions constructed analytically in Sec.~\ref{subsec:Langer_approx} provide a fast and reliable estimate of the eigenfunctions and eigenvalues, which can be used to study quantitatively the soliton core properties.

\subsection{Analytical estimation of soliton core}
\label{subsec:core-structure}

In this subsection, before addressing the core-halo relations, we shall compare the ground-state wave function ($n=1$) with the soliton density profile found in numerical simulations. 

It has been found in numerical simulations that the central core structure of FDM halos is well described by the following fitting form \cite{Schive_etal2014a,Schive_etal2014b}: 
\begin{align}
 \rho_{\rm soliton}(r)=\frac{\rhoc}{\{1+\gamma(r/\rc)^2\}^8}, 
\label{eq:fitting_profile}
\end{align}
with the central density $\rhoc$ given by\footnote{Here, the quantities $\rhoc$  and $\rc$ are defined as the comoving density and comoving radius, respectively.}
\begin{align}
 \rhoc = \frac{0.019}{a}\,\Bigl(\frac{m_\phi}{10^{-22}}\Bigr)^{-2}\Bigl(\frac{\rc}{1\,\mbox{kpc}}\Bigr)^{-4}\,\,[M_\odot\mbox{pc}^{-3}].
\label{eq:rhoc_Schive}
\end{align}
In the above, the core size $\xc$ is defined to be the radius at which the density drops to one-half of the central density, i.e., $\rho_{\rm soliton}(\xc)=\rhoc/2$. This gives the constant $\gamma =2^{1/8}-1\simeq 0.091$.

Similarly, as we have seen in Fig.~\ref{fig:eigen_func_log_alpha10_100_1000}, the ground-state eigenfunctions ($n=1$), depicted as black curves, commonly have a flat core followed by a sharp drop, irrespective of the parameter $\alpha$. Although our treatment ignoring the self-gravity cannot precisely predict the soliton core density $\rho_{\rm c}$, making use of the asymptotic properties of the Airy function, one can analytically express the size of the soliton core structure near the origin. From Eq.~(\ref{eq:eigen_func_analytical_ell0}), we have 
\begin{align}
u_{n}(x)&=\frac{\tilde{u}_{n}(x)}{x} 
\nonumber
\\
&\stackrel{x\ll1}{\longrightarrow}
\frac{1}{\sqrt{\pi}\,\bigl\{g_0(x)\bigr\}^{1/4}}\,\sin\Bigl[\frac{2}{3}\bigl\{-z(x) \bigr\}^{3/2}+\frac{\pi}{4}\Bigr]
\nonumber
\\
&\simeq
(-1)^{n+1}\frac{\alpha^{1/2}\{g_0(0)\}^{1/4}}{\pi^{1/2}}\,\Bigl(1-\beta\,x^2+\cdots\Bigr),
\label{eq:wave_func_near_origin}
\end{align}
where the function $z$ is Taylor expanded in the last line. Rewriting further the derivatives of $z$ in terms of those of the function $g_0$, the coefficient $\beta$ is expressed with a help of Eqs.~(\ref{eq:choice_p}) and (\ref{eq:p_z_func_x_minus}) as 
\begin{align}
 \beta&= \frac{\alpha}{6}\,g_0(0) + \frac{5}{96}\Bigl\{\frac{g_0'(0)}{g_0(0)}\Bigr\}^2-\frac{1}{24}\frac{g_0''(0)}{g_0(0)}.
\label{eq:beta_general}
\end{align}
Note that in deriving Eq.~(\ref{eq:beta_general}), we do not use any functional form of $g_0$. To get a more explicit expression, we substitute the function $g_0(x)$ given at Eq.~(\ref{eq:SP_eq3_ell0}). We then obtain
\begin{align}
\beta=\frac{\alpha}{6}\Bigl(1+\frac{\mathcal{E}}{\alpha}\Bigr)+\frac{17+32\mathcal{E}/\alpha}{1152(1+\mathcal{E}/\alpha)^2}.
\label{eq:beta_NFW}
\end{align}
Equating especially the ground-state eigenfunction $|u_1(x)|^2/|u_1(0)|^2$ with $\rho_{\rm soliton}(x)/\rho_{\rm c}$ at Eq.~(\ref{eq:fitting_profile}), we find that the dimensionless core radius $\xc$, defined by $\xc\equiv\rc/\rs$, is expressed as
\begin{align}
 \xc = p \,\beta^{-1/2} = p \,\sqrt{\frac{6}{\alpha(1+\mathcal{E}/\alpha)}},
\label{eq:x_core_ell_zero}
\end{align}
with the constant $p$ chosen to be $p=0.65$. Here, we have ignored the second term in Eq.~(\ref{eq:beta_NFW}), which merely gives a subdominant contribution. To be precise, in Eq.~(\ref{eq:x_core_ell_zero}), a relevant numerical factor is $p=\sqrt{4\gamma}\simeq0.60$, but we here adopt the valpue of $0.65$, which in fact gives a better agreement with Eq.~(\ref{eq:x_core_ell_zero}) when comparing the analytically estimated density profile with the fitting formula below.

\begin{figure*}[tb]
 \includegraphics[width=18cm,angle=0]{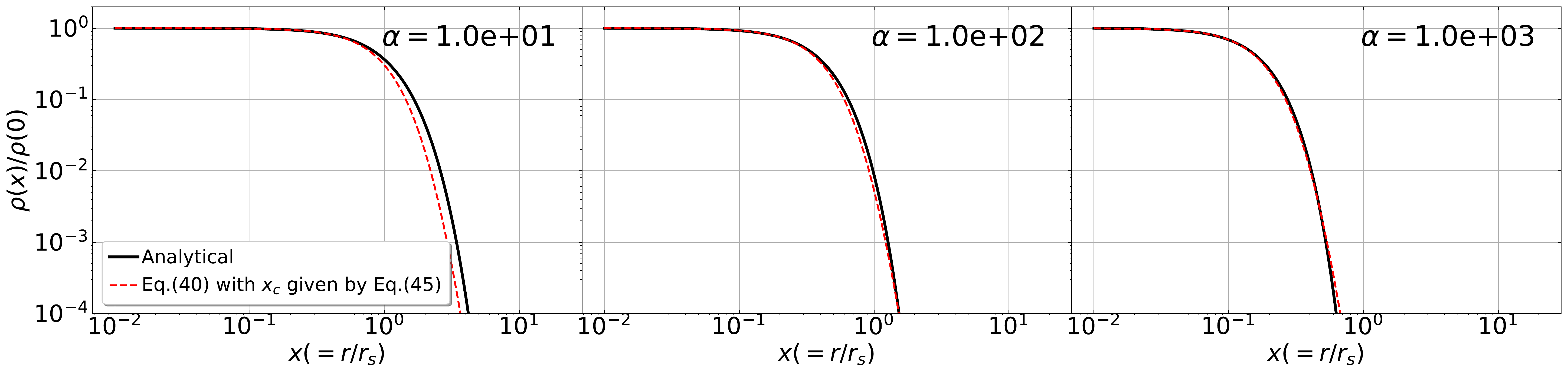}
\hspace*{0.5cm}
\caption{Comparison of the ground-state $(n=1)$ density profile (black solid) with the fitting function of the soliton density profile (red dashed). Plotted results are the normalized density profile, $\rho(x)/\rho(0)$, so that it approaches unity at $x\to0$. From left to right panels, the results for $\alpha=5\times10^3$, $10^4$, and $5\times10^4$ are shown. In plotting the fitting function in Eq.~(\ref{eq:fitting_profile}), we adopt the core radius $\xc$ given at Eq.~(\ref{eq:x_core_ell_zero}). Note that the ground-state density profiles are identical to those shown in Fig.~\ref{fig:eigen_func_log_alpha10_100_1000} (see thin black solid lines).
\label{fig:comparison_profile}
}
\end{figure*}

Using the expression of the core radius in Eq.~(\ref{eq:x_core_ell_zero}), Fig.~\ref{fig:comparison_profile} compares the normalized density profile obtained from the analytical expression of the ground-state wave function with Eq.~(\ref{eq:fitting_profile}). Plugging Eq.~(\ref{eq:x_core_ell_zero}) into Eq.~(\ref{eq:fitting_profile}), the analytical expression depicted as solid lines describes remarkably well the fitting function (red dashed) over a wide range of the parameter $\alpha$ even at the outskirt of the profile. Thus, our analytical result can give a good description for the simulated soliton profile. In particular, Eq.~(\ref{eq:x_core_ell_zero}) is the key to derive analytically the core-halo relation, which we will discuss in more detail.

\section{Predicting soliton core-halo relations}
\label{sec:result}

In this section, on the basis of the analytical expression at Eq.~(\ref{eq:x_core_ell_zero}), we investigate in detail properties of the soliton core, focusing particularly on its relation to the halo mass and other parameters.

\subsection{Core-halo mass-concentration relations}
\label{subsec:core-halo_mass-concentration_relation}

Let us recast the expression at Eq.~(\ref{eq:x_core_ell_zero}) with the {\it comoving} core radius $\rc$, showing explicitly the parameter dependence. Through the relation $\rc=\xc\,\rs$ with the scale radius given by $\rs=r_{\rm vir}/\cvir$, using the expression of $\alpha$ at Eq.~(\ref{eq:alpha_value}) and the definition of $r_{\rm vir}$ at Eq.~(\ref{eq:virial_radius}) gives
\begin{widetext}
\begin{align}
\rc&=p \sqrt{\frac{6}{1+\mathcal{E}/\alpha}} \frac{r_{\rm vir}/\cvir}{\alpha^{1/2}}
\nonumber
\\
&\simeq 1.83 \,\,\mbox{[kpc]}\,\,a^{-1/2}\,\Bigl(\frac{p}{0.65}\Bigr)\,\Bigl(\frac{\mass}{10^{-22}{\rm eV}}\Bigr)^{-1}\Bigl(\frac{M_{\rm h}}{10^9\,M_\odot}\Bigr)^{-1/3}\Bigl(\frac{\Delta_{\rm vir}}{200}\Bigr)^{-1/6}\Bigl(\frac{\Omega_{\rm m,0}h^2}{0.147}\Bigr)^{-1/6}\,
\Bigl\{\frac{g(\cvir)}{1+\mathcal{E}/\alpha}\Bigr\}^{1/2},
\label{eq:r_core_halo_mass}
\end{align}
\end{widetext}
where the function $g(\cvir)$ is defined in Eq.~(\ref{eq:def_g_func}).

The above relation is compared to Eq.~(7) of Ref.~\cite{Schive_etal2014b}, who derived it based on the kinematic argument with a help of numerical experiments. Apart from the last factor, the dependencies on FDM mass, halo mass, and cosmological parameters exactly coincide with each other\footnote{In Eq.~(7) of Ref.~\cite{Schive_etal2014b}, the cosmological parameters are set to typical values, and their expression of the core radius apparently misses the dependence on $\Omega_{\rm m,0}h^2$. Note also that Eq.~(7) is given as a physical radius, and thus the scale factor dependence differs from ours by the factor of $a$. }. Also, the numerical prefactor of $1.83$ in our expression is rather close to theirs.

However, an important difference is the last factor,  $\{g(\cvir)/(1+\mathcal{E}/\alpha)\}^{1/2}$. While the numerator of this factor is given as a function of $\cvir$, the denominator, $1+\mathcal{E}/\alpha$, depends on the parameter $\alpha$. Thus, through Eq.~(\ref{eq:alpha_value}), the actual parameter dependence of $\rc$ can differ significantly from the one in Ref.~\cite{Schive_etal2014b}. 

To see the explicit dependence of the expression (\ref{eq:r_core_halo_mass}) on  the halo properties, we hereafter fix the cosmological parameters and FDM mass, respectively, to $\Omega_{\rm m,0}=0.276$, $h=0.677$, $\Omega_\Lambda=0.724$, and $m_\phi=8\times10^{-23}$\,eV. Then, for a given redshift $z$, the soliton core size $\rc$ is described as a function of the halo mass and concentration parameter, $M_{\rm h}$ and $\cvir$. While the concentration parameter is also given as a function of the halo mass and redshift, its dependence is known to be sensitive to the cosmology, mainly through the initial power spectrum. Hence, as a generic prediction of the core-halo relation, we first consider $\cvir$ to be an independent parameter of the halo mass. We shall then introduce the models of concentration-halo mass relation, and discuss their core-halo relations (Sec.~\ref{subsec:core-halo_mass_relations}).

\begin{figure*}[tb]
\begin{center}
\includegraphics[width=18.0cm,angle=0]{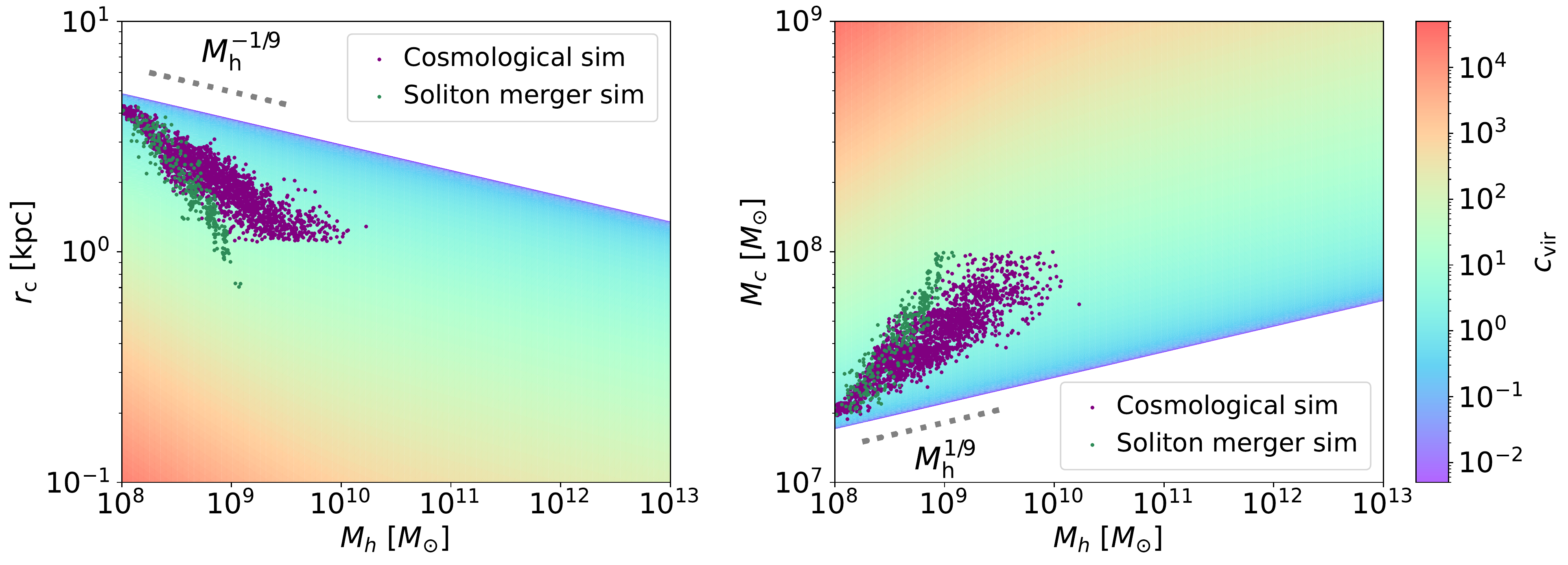}
\end{center}
\caption{Soliton core-halo relation given as function of halo mass $M_{\rm h}$ and concentration parameter $\cvir$, with the latter dependence shown in the color scales. The analytical predictions of the core radius vs halo mass (left) and core mass vs halo mass (right) relations, given respectively at Eqs.~(\ref{eq:r_core_halo_mass}) and (\ref{eq:M_core_halo_mass}), are plotted at $z=0$. Here, we adopt the mass of the FDM and the cosmological parameters as follows: $m_\phi=8\times10^{-23}$\,eV, $\Omega_{\rm m,0}=0.276$, $\Omega_\Lambda=0.724$ and $h=0.677$. For comparison, we also plot the results measured from the numerical simulations of S-P equation. The dark green and purple points respectively represent the results from soliton merger simulations \cite{Jowett_etal2022} and cosmological simulations \cite{May_Springel2021}. Note that in each panel, regions with no color indicate that soliton core is not theoretically allowed to form. The boundary on the allowed core radius and/or core mass is described by a single power-law function of halo mass, whose scaling is depicted as gray dashed lines (see the text in detail). 
\label{fig:core_halo_mass_concentration}
}
\end{figure*}
Left panel of Fig.~\ref{fig:core_halo_mass_concentration} plots the predicted core radius given as a function of $M_{\rm h}$ and $\cvir$, with the latter dependence shown in the color scales. In similar manner, right panel plots the predicted core mass. Here, we adopt the fitting form of the soliton density profile at Eq.~(\ref{eq:fitting_profile}) with the soliton core density $\rho_{\rm c}$ given by Eq.~(\ref{eq:rhoc_Schive}). The core mass is then estimated from (e.g., \cite{Nori_Baldi2021}):
\begin{align}
 M_{\rm c}\simeq 4\pi(0.2225)\rho_{\rm c}\,\rc^3,
\label{eq:soliton_core_mass}
\end{align}
which gives 
\begin{widetext}
\begin{align}
 M_{\rm c}= 2.91\times 10^7\,\,
[M_\odot]\,\,a^{-1/2}\,\Bigl(\frac{p}{0.65}\Bigr)^{-1}\,\Bigl(\frac{\mass}{10^{-22}{\rm eV}}\Bigr)\Bigl(\frac{M_{\rm vir}}{10^9\,M_\odot}\Bigr)^{1/3}\Bigl(\frac{\Delta_{\rm vir}}{200}\Bigr)^{1/6}\Bigl(\frac{\Omega_{\rm m,0}h^2}{0.147}\Bigr)^{1/6}
\Bigl\{\frac{g(\cvir)}{1+\mathcal{E}_n/\alpha}\Bigr\}^{-1/2}.
\label{eq:M_core_halo_mass}
\end{align}
\end{widetext}
Since the core mass shown here partly uses the relation determined by the numerical simulations [in particular, the core density at Eq.~(\ref{eq:rhoc_Schive})], Eq.~(\ref{eq:M_core_halo_mass}) is not strictly the first-principle prediction, compared to the core radius shown in the left panel, where we only used the analytical results given in Sec.~\ref{sec:Analytic_solutions_ell0}. Nevertheless, the predicted core mass can be used as an independent cross check, and it provides some insights into the one obtained from numerical simulations, as we will discuss below.

In both panels of Fig.~\ref{fig:core_halo_mass_concentration}, we also plot the results measured from numerical simulations, depicted as filled circles. These data are taken from Ref.~\cite{Jowett_etal2022}. Two different colors indicate the results obtained from either soliton merger simulations (dark green) or simulations started from the cosmological initial condition (purple, Ref.~\cite{May_Springel2021})\footnote{To be precise, simulation data of Ref.~\cite{May_Springel2021} have started from the CDM initial condition. Thus, unlike the FDM initial condition, no small-scale cutoff was imposed in the initial conditions. Nevertheless, the evolved power spectra measured at later time are shown to look quantitatively similar to those expected from the FDM initial condition.}. Multiplying by the scale factor $a^{1/2}$, the results are all scaled to those at $z=0$.

In Fig.~\ref{fig:core_halo_mass_concentration}, a cautious remark is that the parameter $\alpha$ becomes smaller than unity at the region satisfying the condition $1\lesssim\,(M_{\rm h}/10^7 M_\odot)^{3/4} \,f(\cvir)$, which roughly corresponds to reddish triangular regions. Although our analytical description of the core-halo relations become inaccurate there, these parameter regions are somewhat extreme and there are in fact no simulation data points. Apart from these points, important findings from Fig.~\ref{fig:core_halo_mass_concentration} are summarized below:
\begin{itemize}
 \item All of the simulation results lie at the regions allowed by predictions. To be precise, measured core radii and core masses are consistent with predictions with moderately small values of concentration parameter, $\cvir\sim\mathcal{O}(1-10)$. 
 \item In predictions, there exist clear boundaries on the allowed core radius and core mass for each halo mass. Denoting, respectively, the upper and lower bounds on the predicted core radius and core mass by $\rc^{\rm limit}$ and $M_{\rm c}^{\rm limit}$, one finds that they gradually changes with halo mass, scaled as $\rc^{\rm limit}\propto M_{\rm h}^{-1/9}$ and $M_{\rm c}^{\rm limit}\propto M_{\rm h}^{1/9}$ (see gray dashed lines). Both of the bounds appear in the limit of $\cvir\to0$. 
\end{itemize}

Making use of the analytical properties of the wave functions in Sec.~\ref{sec:Analytic_solutions_ell0}, we can derive explicit expressions for the critical radius $\rc^{\rm limit}$ and mass $M_{\rm c}^{\rm limit}$ as follows. We first notice that taking the limit $\cvir\to0$, the parameter $\alpha$ yields $\alpha\to\infty$. This implies that we can safely use the approximate expression at Eq.~(\ref{eq:eigenvalue_approximation}) to evaluate the ground-state $(n=1)$ eigenvalue. Then, writing the parameter $\alpha$ given in Eq.~(\ref{eq:alpha_value}) as $\alpha=\tilde{\alpha}\,f(\cvir)$, the factor of $\{g(\cvir)/(1+\mathcal{E}/\alpha)\}^{1/2}$ can be computed explicitly in the limit of $\cvir\to0$. We obtain
\begin{align}
 \Bigl\{\frac{g(\cvir)}{1+\mathcal{E}_n/\alpha}\Bigr\}^{1/2} 
\,\,\stackrel{\cvir\to0}{\longrightarrow}\,\,\tilde{\alpha}^{1/6}\,\Bigl(\frac{8}{9\pi}\Bigr)^{1/3}.
\label{eq:factor_cvir_limit}
\end{align}
Recalling that the quantity $\tilde{\alpha}$ depends on various model parameters including the FDM mass and halo mass [Eq.~(\ref{eq:alpha_value})], plugging Eq.~(\ref{eq:factor_cvir_limit}) into Eq.~(\ref{eq:r_core_halo_mass}) finally yields the following relation:
\begin{widetext}
\begin{align}
\rc^{\rm limit}= 3.59\,[\mbox{kpc}]\,\,a^{-1/3}\,\Bigl(\frac{p}{0.65}\Bigr)\,\Bigl(\frac{m_\phi}{10^{-22}}\Bigr)^{-2/3}\Bigl(\frac{M_{\rm h}}{10^9\,M_\odot}\Bigr)^{-1/9}\Bigl(\frac{\Delta_{\rm vir}}{200}\Bigr)^{-2/9}\Bigl(\frac{\Omega_{\rm m,0}h^2}{0.147}\Bigr)^{-2/9}, 
\label{eq:r_core_upper_limit}
\end{align}
\end{widetext}
which gives the upper bound on the core radius. Similarly, the lower bound on the soliton core mass is obtained by substituting Eq.~(\ref{eq:factor_cvir_limit}) into Eq.~(\ref{eq:M_core_halo_mass}): 
\begin{widetext}
\begin{align}
M_{\rm c}^{\rm limit}= 1.48\times10^7\,[M_\odot]\,\,a^{-2/3}\,\Bigl(\frac{m_\phi}{10^{-22}}\Bigr)^{2/3}\Bigl(\frac{M_{\rm h}}{10^9\,M_\odot}\Bigr)^{1/9}\Bigl(\frac{\Delta_{\rm vir}}{200}\Bigr)^{2/9}\Bigl(\frac{\Omega_{\rm m,0}h^2}{0.147}\Bigr)^{2/9}.
\label{eq:M_core_lower_limit}
\end{align}
\end{widetext}

Equations~(\ref{eq:r_core_upper_limit}) and (\ref{eq:M_core_lower_limit}) derived above precisely reproduce the boundary curves shown in Fig.~\ref{fig:core_halo_mass_concentration}. The presence of upper (lower) bound on core radius (mass) are related to the structure of halo potential near the center. In other words, the inner structure of halo profile determines the allowable soliton core size and mass, and they can be indeed changed depending on the inner slope of halo density profile. In Appendix \ref{sec:soliton_core_cvir_limit}, we examine this in detail in a class of generalized halo profiles, and show that the upper bound on the core radius $\rc^{\rm limit}$ generally scales with respect to the inner slope of the density profile $s$, defined by $\rho\propto r^{-s}$ $(0\leq s <2)$,  as [Eq.~(\ref{eq:rcore_limit_general})]:
\begin{widetext}
\begin{align}
 \rc^{\rm limit}=3.59\,\,\mbox{[kpc]}\,\,\frac{\mathcal{T}(s)}{\mathcal{T}(1)}
a^{-1/(4-s)}\,\Bigl(\frac{p}{0.65}\Bigr)\,\Bigl(\frac{m_\phi}{10^{-22}\,\mbox{eV}}\Bigr)^{-2/(4-s)}\Bigl(\frac{M_{\rm h}}{10^9\,M_\odot}\Bigr)^{-s/3/(4-s)}
\Bigl(\frac{\Delta_{\rm vir}}{200}\,\,\frac{\Omega_{\rm m,0}h^2}{0.147}\Bigr)^{-(1-s/3)/(4-s)},
\nonumber
\end{align}
\end{widetext}
where the function $\mathcal{T}$, defined at Eq.~(\ref{eq:def_func_T}), is a monotonically decreasing function of the slope $s$, and it becomes zero when the slope approaches $2$. Although the above expression is valid at $0\leq s<2$, the behavior of the function $\mathcal{T}$ suggests that the FDM soliton, if exists, does not form a flat core for a halo having a steeper inner slope of $s\geq2$.

\subsection{Core-halo mass relations in CDM and FDM cosmologies}
\label{subsec:core-halo_mass_relations}

In this subsection, incorporating the concentration-mass (C-M) relation $\cvir(M_{\rm h})$ into the results shown Fig.~\ref{fig:core_halo_mass_concentration}, we compare the predicted core-halo relations with simulation results.

\begin{figure}[tb]
 \includegraphics[width=8.5cm,angle=0]{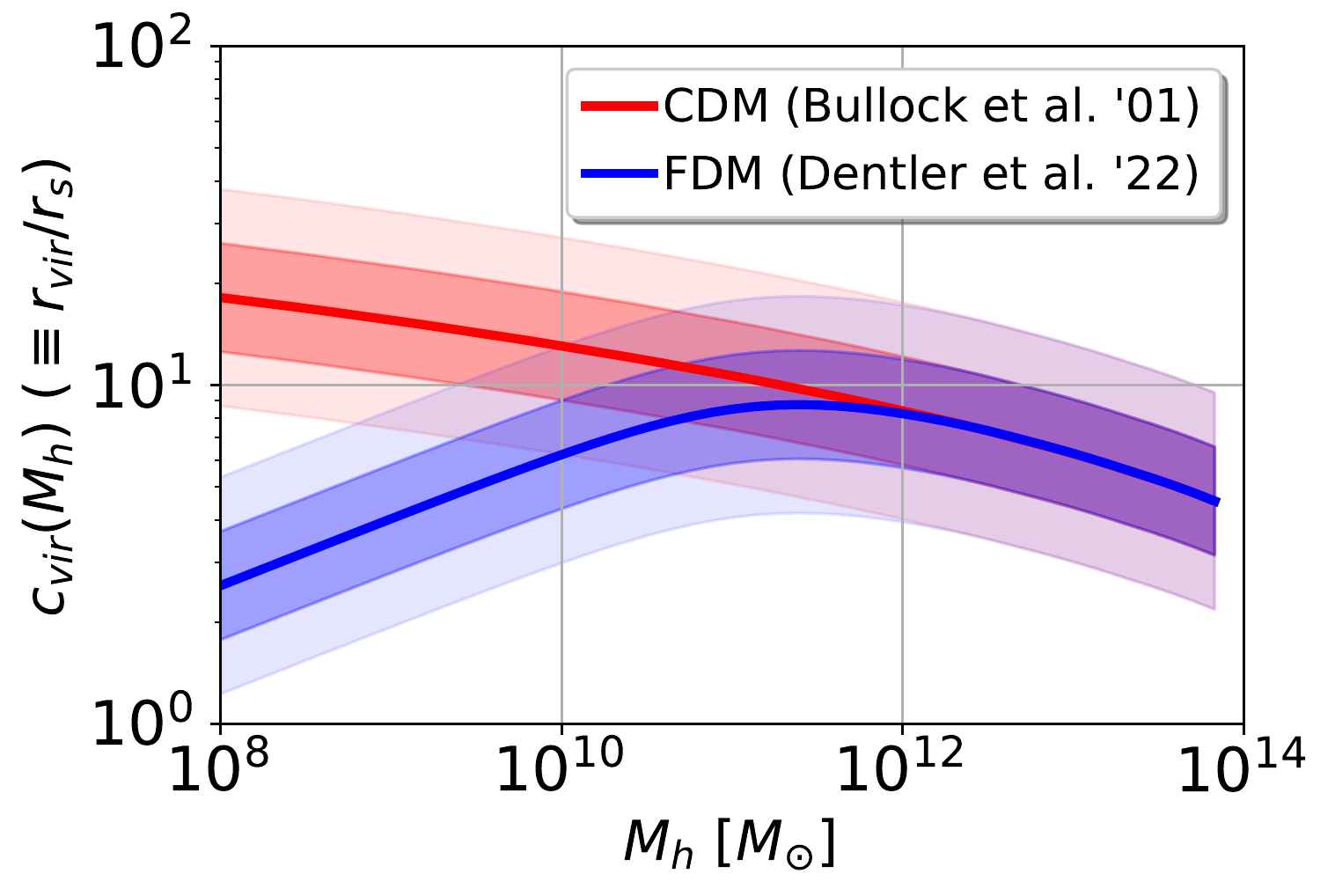}
\caption{Concentration-mass (C-M) relation for the FDM and CDM models. Adopting the analytical models presented in Refs.~\cite{Dentler_etal2022,Bullock_etal2001} (see Appendix \ref{subsec:C-M_relations} for their analytical expressions),  the concentration parameter $\cvir$ is plotted as a function of halo mass in blue and red lines for FDM and CDM models, respectively. The faint and dark shaded area indicate the $1\sigma$ and $2\sigma$ errors, assuming the log-normal distribution of $\cvir$ with dispersion given by $0.16$\,dex. 
\label{fig:cM_relation}
}
\end{figure}
It is known that the C-M relation acquires various dependencies on cosmology and redshift through the formation and merger history of halos. As a result, at low-mass halos of our interest, it becomes rather sensitive to the cutoff of the initial power spectrum. Here, as representative examples, we examine the two C-M relations given for different cosmological models. One is the C-M relation of the FDM model. The linear power spectrum of this model exhibits a sharp cutoff at the wave number $k_{1/2}\sim4.5 (m_\phi/10^{-22}\,\mbox{eV})^{4/9}$\,Mpc$^{-1}$ \cite{Hu_Barkana_Gruzinov2000,Marsh_review2016}, and this cutoff leads to the formation of low-concentration halos with a suppressed abundance, analogous to the case in the warm dark matter model (e.g., Refs.~\cite{Schneider_etal2012,Ludlow_etal2016}). At present, there is a little work to numerical study the C-M relation in a relevant cosmological setup. We adopt the analytical C-M relation proposed by Ref.~\cite{Dentler_etal2022}, who applied it to predict the two-point statistics of galaxies and weak lensing based on the halo model prescription. Another C-M relation we examine is the one of the CDM model. In contrast to the FDM model, the power spectrum of the CDM model does not have any typical cutoff at relevant scales of structure formation. Thus, a sizable amount of halos is formed, and this results in the high concentration halos at small halo masses. We adopt the C-M relation given by Ref.~\cite{Bullock_etal2001}, which has been calibrated by cosmological $N$-body simulations to quantitatively match the measured C-M relation (see also Refs.~\cite{Diemer_Joyce2019,Wang_etal2020,Ishiyama_etal2021} for recent improved modeling).

The analytical expressions for the C-M relation of both models are summarized in Appendix \ref{subsec:C-M_relations}. Based on these, Fig.~\ref{fig:cM_relation} shows the C-M relations of the FDM (blue) and CDM (red) models. Since the C-M relation measured in simulations is known to have a large scatter (e.g., Refs.~\cite{Jing2000,Bullock_etal2001,Neto_etal2007,Reed_etal2011,Bhattacharya_etal2013,Diemer_Kravtsov2015}), we also show in Fig.~\ref{fig:cM_relation} the $1$ and 2$\sigma$ errors around the mean C-M relation, assuming the log-normal distribution with the scatter of $0.16$ dex \cite{Diemer_Kravtsov2015}.

\begin{figure*}[tb]
\begin{center}
\includegraphics[width=18.0cm,angle=0]{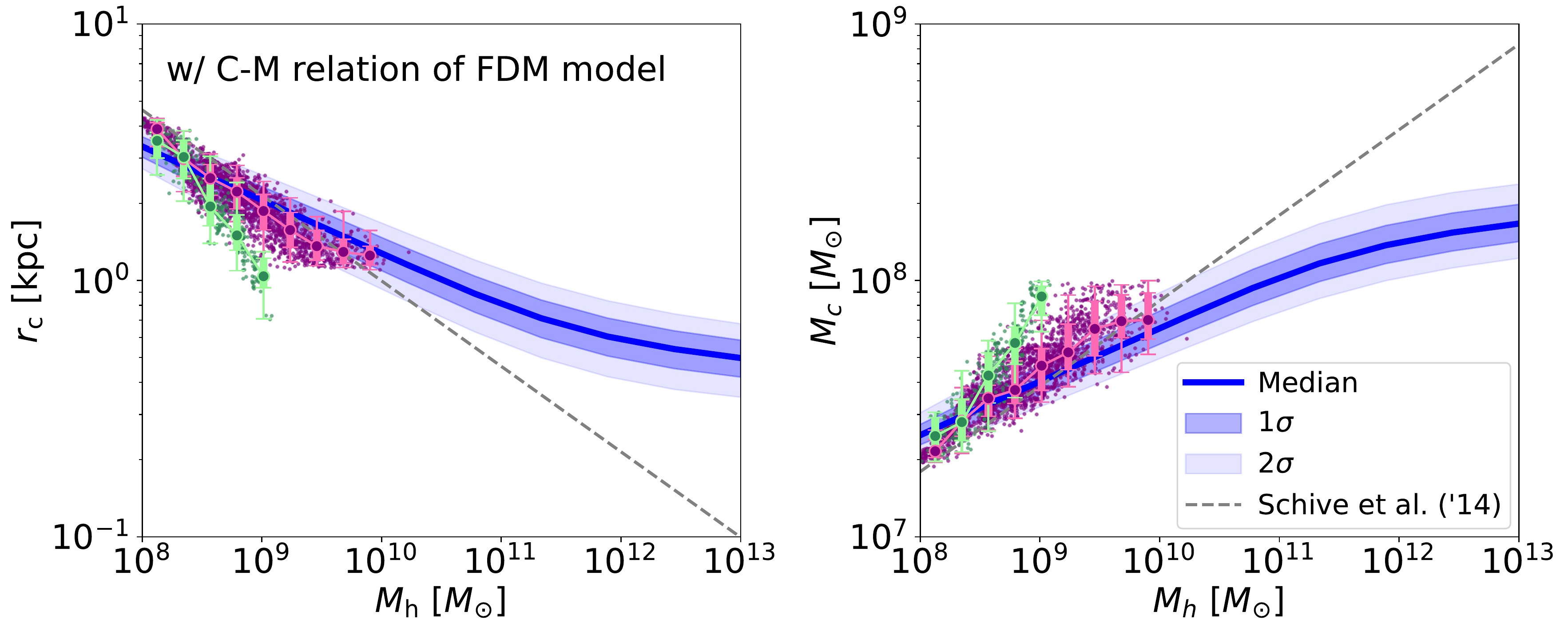}
\end{center}
\caption{The soliton core-halo mass relation adopting the C-M relation for the FDM model. Incorporating the C-M relation of Ref.~\cite{Dentler_etal2022} into the predictions shown in Fig.~\ref{fig:core_halo_mass_concentration}, the results of the core radius vs halo mass (left) and the core mass vs halo mass (right) relations are plotted, together with measured results from numerical simulations, for which the median values and dispersions are also evaluated in each halo mass bin and are plotted as large filled circles and error bars, respectively . In plotting the predictions, we assume, for each halo mass, the log-normal distribution of $\cvir$, and evaluate the median and the scatter in the core-halo relation. In each panel, the thick solid line is the median relation, while the faint and dark shaded areas respectively show the $1\sigma$ and $2\sigma$ scatter arising from the scatter in $\cvir$. Note that the median relations shown here are hardly distinguishable from the predictions computed with the mean C-M relation. For reference, the gray dashed lines are the scaling relations numerically found by Ref.~\cite{Schive_etal2014b}. 
\label{fig:core_halo_mass_FDM}
}
%
\begin{center}
\includegraphics[width=18.0cm,angle=0]{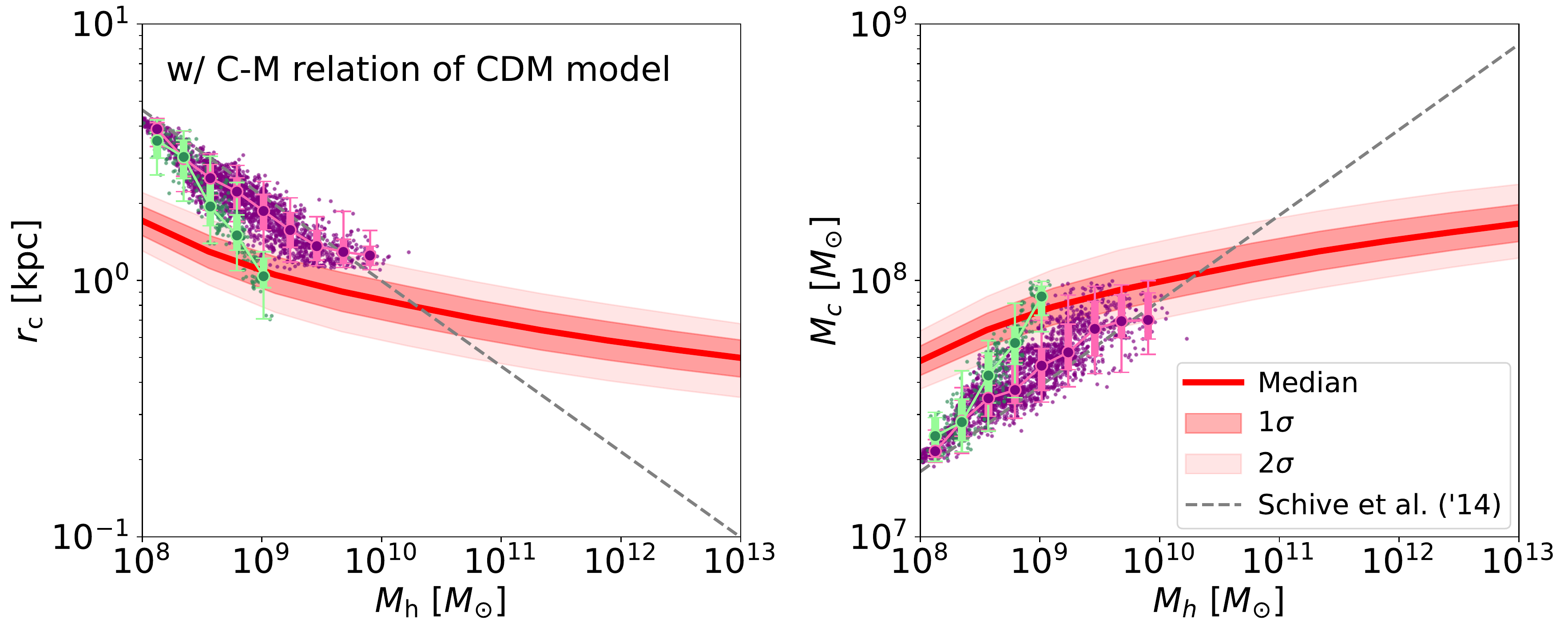}
\end{center}
\caption{Same as Fig.~\ref{fig:core_halo_mass_FDM}, but the predictions adopting the C-M relation of the CDM model by Ref.~\cite{Bullock_etal2001} are shown. 
\label{fig:core_halo_mass_CDM}
}
\end{figure*}

Using these C-M relations and their scatter in Fig.~\ref{fig:cM_relation}, we compute the core-halo relations, and the predictions are plotted in Figs.~\ref{fig:core_halo_mass_FDM} and \ref{fig:core_halo_mass_CDM} for FDM and CDM models, respectively. Together with the fitting function found numerically by Ref.~\cite{Schive_etal2014b} (gray dashed), simulation results are also shown as small filled circles, for which we further estimate the median values and dispersions, depicted, respectively, as large filled circles and error bars.

Overall, the predicted core-halo relations exhibit a non-power-law behavior for both the FDM and CDM models. At the small halo masses of $M_{\rm h}\lesssim10^{11}\,M_\odot$, differences between the predictions become manifest. Obviously, the results from the C-M relation of the CDM model, which predicts high-concentration halos, fails to reproduce the trend seen in numerical simulations. On the other hand, adopting the C-M relation of the FDM model, the predicted core-halo relations gives a close agreement with the results obtained from numerical simulations. Interestingly, at $M_{\rm h}\sim10^8-10^9\,M_\odot$, the predicted core-halo relations get closer to the scaling relation found by Ref.~\cite{Schive_etal2014b} (gray dashed), which predicts the core-halo relation in a power-law form of $r_{\rm c}\propto M_{\rm h}^{-1/3}$ and $M_{\rm c}\propto M_{\rm h}^{1/3}$. Also, scatters in the predicted core-halo relations resemble those shown in the simulations. 

These results suggest that provided the C-M relation well-calibrated with numerical simulations, our analytical formulas given at Eqs.~(\ref{eq:r_core_halo_mass}) and (\ref{eq:M_core_halo_mass}) can successfully describe the core-halo relations found in numerical simulations. Nevertheless, one caveat to be noted is that the data points shown in Figs.~\ref{fig:core_halo_mass_FDM} and \ref{fig:core_halo_mass_CDM} are obtained from the simulations with a small box size, $L=10\,h^{-1}$\,Mpc for the cosmological simulations \cite{May_Springel2021}, and $L=300$\,kpc for the soliton merger simulations. Further, while the latter simulations are not strictly made with a cosmological setup, the initial conditions of the former simulations is not precisely consistent with the FDM model having a small-scale cutoff. In this respect, in both simulations, the evolved halos may not necessarily trace the C-M relations expected from those obtained from a relevant cosmological setup. Indeed, we see a small discrepancy with predictions, which appears manifest around the halo mass of $M_{\rm h}\sim10^9\,M_\odot$. In addition, the analytical C-M relation for the FDM model is not designed to account for the low-mass halos considered here. In this respect, the predicted core-halo relations adopting the C-M relation of Ref.~\cite{Dentler_etal2022} might not be accurate at $M_{\rm h}\lesssim10^{11}M_\odot$. Still, the results shown in Fig.~\ref{fig:core_halo_mass_FDM} is very promising, and a more quantitative investigation along the direction would shed light on clarifying the origin and diversity of core-halo relations.

Finally, a closer look at the predictions in the FDM model reveals that the scatter in core-halo relations gets smaller as decreasing the halo mass. This is because the FDM model prefers low-concentration halos at small mass scales. As we have seen in Sec.~\ref{subsec:core-halo_mass-concentration_relation}, the allowable size and mass of the soliton core become converged to a certain value for each halo mass in the limit of $\cvir\to0$ [see Eqs.~(\ref{eq:r_core_upper_limit}) and (\ref{eq:M_core_lower_limit})]. As a result, the scatter in the predicted core-halo relation gets reduced for low-concentration halos. Indeed, similar trend can be clearly seen in numerical simulations, and data points from the two different simulations converge to a similar core radius and core mass at $M_{\rm h}\sim 10^8\,M_\odot$, giving consistent results with predictions of the FDM model.

\section{Discussion: impact of soliton self-gravity}
\label{sec:discussion}

So far, we have ignored the self gravity of soliton core, and constructed analytically the eigenfunctions of the S-P system under the potential of a spherically symmetric halo. Focusing particularly on the ground-state eigenfunction, we derived the analytical expression of the soliton core, which is then used to predict the soliton core-halo relations. In this section, we discuss the effect of soliton self-gravity, and estimate in particular its impact on the soliton core radius.

As we have seen in Sec.~\ref{subsec:S-P_eq}, a proper account of the soliton self-gravity is to add the contribution of the soliton core to the gravitational potential, on top of the potential of the background halo. The effective potential of Eq.~(\ref{eq:stationary-SP_eq2}) is then changed to
\begin{align}
&-\alpha\frac{\log(1+x)}{x}+\frac{\ell(\ell+1)}{x^2}
\nonumber
\\
&\qquad\longrightarrow\,\,
-\alpha\frac{\log(1+x)}{x}+\frac{\ell(\ell+1)}{x^2}
+\Delta V(x),
\label{eq:potential_SP_eq_with_soliton}
\end{align}
where the correction term $\Delta V$ is related to the gravitational potential of the soliton, $\Psi_{\rm soliton}$: 
\begin{align}
\Delta V(x)=\frac{2\mass^2\rs^2a^2}{\hbar^2}\Psi_{\rm soliton}(x).
\label{eq:Delta_V}
\end{align}

Since the potential $\Psi_{\rm soliton}$ is determined by the Poisson equation sourced by the square of the wave function itself [see Eqs.~(\ref{eq:Poisson_eq}) with (\ref{eq:def_delta_rho_phi})], the S-P equation having the effective potential at Eq.~(\ref{eq:potential_SP_eq_with_soliton}) is no longer reduced to a linear eigenvalue problem. Nevertheless, if the potential of the self-gravitating soliton is small enough compared to the background halo potential, a perturbative treatment can be applied, and the impact of soliton self-gravity is estimated based on our treatment in Sec.~\ref{sec:Analytic_solutions_ell0}.

Let us suppose that the soliton core is described by the ground-state eigenfunction in Sec.~\ref{subsec:Langer_approx}, with its density profile approximately 
given by the fitting function at Eq.~(\ref{eq:fitting_profile}). Adopting this fitting form, the Poisson equation is analytically solved, and the gravitational potential induced by the soliton core is expressed as \cite{Chiang_etal2021}
\begin{widetext}
\begin{align}
 \Psi_{\rm soliton}(x)&= -\pi\,G\,\frac{\rhoc\,\rs^2}{a}\,\frac{\xc^2}{53760\,\gamma^{3/2}}\,
\Bigl[\frac{\gamma^{1/2}\,h(x/\xc)}{[1+\gamma(x/\xc)^2]^6}+\frac{3465}{x/\xc}\tan^{-1}\Bigl(\frac{\gamma^{1/2}x}{\xc}\Bigr)\Bigr]
\label{eq:potential_FDM}
\end{align}
\end{widetext}
with $\gamma\simeq0.091$ and the function $h(y)$ given by
\begin{align}
 h(y)&=3465\gamma^5 y^{10}+19635\gamma^4 y^8 + 45738\gamma^3y^6 
\nonumber
\\
&+ 55638\gamma^2y^4 + 36685 \gamma y^2 + 11895.
\label{eq:def_of_func_h}
\end{align}
Note that the quantity $x$ is the dimensionless radius normalized by $\rs$, $x=r/\rs$. 

Plugging the explicit functional form into Eq.~(\ref{eq:potential_SP_eq_with_soliton}), we can repeat the same analysis as we have done in Sec.~\ref{sec:Analytic_solutions_ell0} as follows. First notice that the function $g_0(x)$ introduced at Eq.~(\ref{eq:SP_eq3_ell0}) is changed to
\begin{align}
 g_0(x) \longrightarrow \frac{\log(1+x)}{x}+\frac{\mathcal{E}}{\alpha} - \frac{\rhoc}{\rhos}\,\tilde{\Psi}_{\rm soliton}(x),
\label{eq:func_g0_with_soliton}
\end{align}
where the function $\tilde{\Psi}$ is related to the potential $\Psi_{\rm soliton}$ through $\Psi_{\rm soliton}=4\pi\,G\,\rho_{\rm c}\rs^2\,\tilde{\Psi}_{\rm soliton}$. One important point to note is that the modification due to soliton potential does not affect the monotonicity of the function $g_0$. That is, the function $g_0$ still has a single turning point, though its location is perturbatively shifted to an inner radius. Thus, even if the soliton self-gravity is taken into account, the ground-state eigenfunction which describes the soliton core structure is still given by Eq.~(\ref{eq:eigen_func_analytical_ell0}), with the function $g_0$ replaced with Eq.(\ref{eq:func_g0_with_soliton}). Hence, taking the limit of $x\ll1$, we have the same asymptotic form of the eigenfunction as we obtained in Sec.~\ref{subsec:core-structure} [see Eq.~(\ref{eq:beta_general})], but the explicit form of the coefficient $\beta$ is now changed to 
\begin{align}
 \beta &\longrightarrow \frac{\alpha}{6}\Bigl\{1+\frac{\mathcal{E}}{\alpha}-\frac{\rho_{\rm c}}{\rho_{\rm s}}\tilde{\Psi}_{\rm soliton}(0)\Bigr\},
\label{eq:beta_with_soliton}
\end{align}
where we have only taken into account the leading-order term proportional to the parameter $\alpha$. This modification implies that the soliton self-gravity can change the core size $\xc$ from Eq.~(\ref{eq:x_core_ell_zero}) to 
\begin{align}
 \xc \longrightarrow p\,\Bigl[\frac{6}{\alpha\,\bigl\{1+\mathcal{E}_1/\alpha+(\rhoc/\rhos)(\xc^2/14/\gamma)\bigr\}}\Bigr]^{1/2},
\label{eq:x_core_with_soliton}
\end{align}
with $p$ given by $0.65$. Here we used the fact that $\tilde{\Psi}_{\rm soliton}(0)=-\xc^2/(14\,\gamma)$. In the above, the energy eigenvalue $\mathcal{E}$ has to be consistently evaluated, taking the soliton potential into account. Employing the perturbative calculation familiar in quantum mechanics, this is estimated to give 
\begin{align}
 \mathcal{E}_1/\alpha &\simeq \mathcal{E}_1^{(0)}/\alpha + \Bigl(\frac{\rhoc}{\rho_{\rm s}}\Bigr) \int dx\, \tilde{u}_1^{(0)}(x)\,\widetilde{\Psi}_{\rm soliton}(x)\,\tilde{u}_1^{(0)}(x),
\label{eq:perturbed_energy_shift}
\end{align}
where the quantity $\mathcal{E}_1^{(0)}$ stands for the unperturbed energy eigenvalue of the ground state, which is obtained analytically from Eq.~(\ref{eq:eigenvalue_Airy}). The function $\tilde{u}_1^{(0)}$ represents the unperturbed ground-state eigenfunction that is properly normalized.

The new expression of the soliton core radius at Eq.~(\ref{eq:x_core_with_soliton}) involves the soliton core radius $\xc$, which we evaluate with the unperturbed result at Eq.~(\ref{eq:x_core_ell_zero}). As a result, the ratio of the new core radius to the unperturbed counterpart ignoring the soliton self-gravity, which we denote by $\RR$, is expressed as
\begin{widetext}
\begin{align}
 \RR\equiv \frac{\xc({\rm w/\,self\,\, gravity})}{\xc({\rm w/o\,\,self\,\, gravity})}
=\Biggl[1 + \frac{\rhoc/\rhos}{1+\mathcal{E}_1/\alpha}\,\Bigl\{
\frac{\xc^2}{14\,\gamma}+\int dx\,\bigl\{\tilde{u}_1^{(0)}(x)\bigr\}^2\tilde{\Psi}_{\rm soliton}(x)\Bigr\}\Biggr]^{-1/2}.
\end{align}
\end{widetext}
Using the explicit expressions for the quantities $\alpha$, $\xc$, $\rhos$ and adopting the relation for $\rhoc$, respectively given at Eqs.~(\ref{eq:alpha_value}), (\ref{eq:x_core_ell_zero}), (\ref{eq:def_of_rhos}) and (\ref{eq:rhoc_Schive}), the above expression is simplified to give
\begin{align}
 \RR=\Bigl[1+1.73
\Bigl\{1-\int dx\, \bigl\{\tilde{u}_1^{(0)}(x)\bigr\}^2 V_{\rm soliton}(x) 
\Bigr\}
\Bigr]^{-1/2},
\label{eq:ratio_core_radius}
\end{align}
where the function $V_{\rm soliton}$ is the reduced (minus) soliton potential defined by $V_{\rm soliton}(x)\equiv -\tilde{\Psi}_{\rm soliton}(x)/\tilde{\Psi}(0)$. Note that the reduced potential $V_{\rm soliton}$ is a monotonically decreasing function of $x$, and it goes to $1$ at $x\to0$ and $0$ at $x\to\infty$. Thus the integral given above generally takes a value in between $0$ and $1$. Accordingly the ratio $\RR$ is expected to be smaller than $1$, and is around $\RR\sim 0.6-1$\footnote{In Eq.~(\ref{eq:ratio_core_radius}), we adopt the soliton core density $\rhoc$ found in numerical simulations, and the depth of the soliton potential relative to the halo potential is fixed. In this sense, there is no control parameter to change the soliton self-gravity itself. Rather, in our case, the soliton self-gravity is determined by the quantities characterizing the halo properties. }.

\begin{figure*}[tb]
\begin{center}
\includegraphics[height=5.5cm,angle=0]{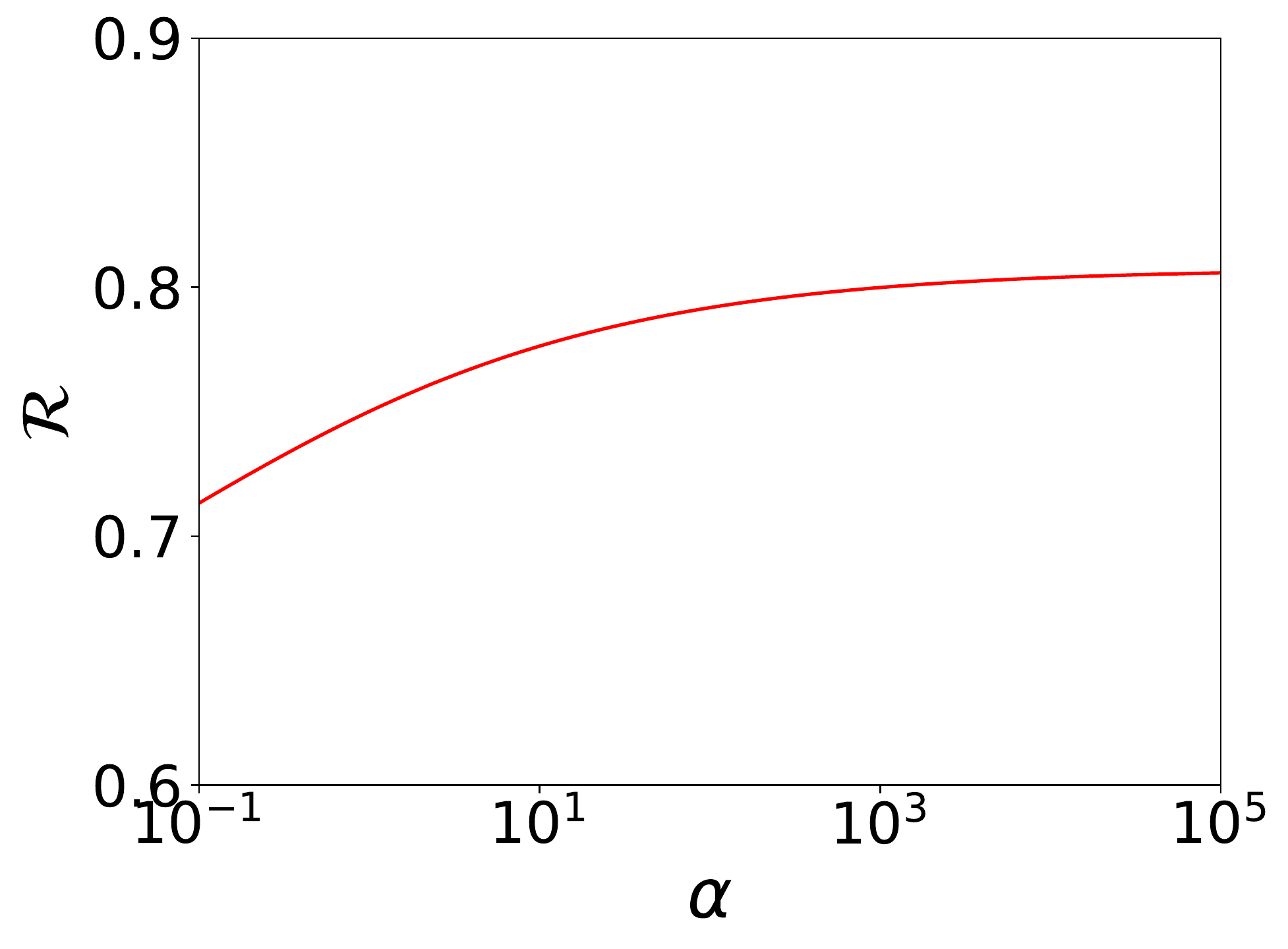}
\includegraphics[height=5.5cm,angle=0]{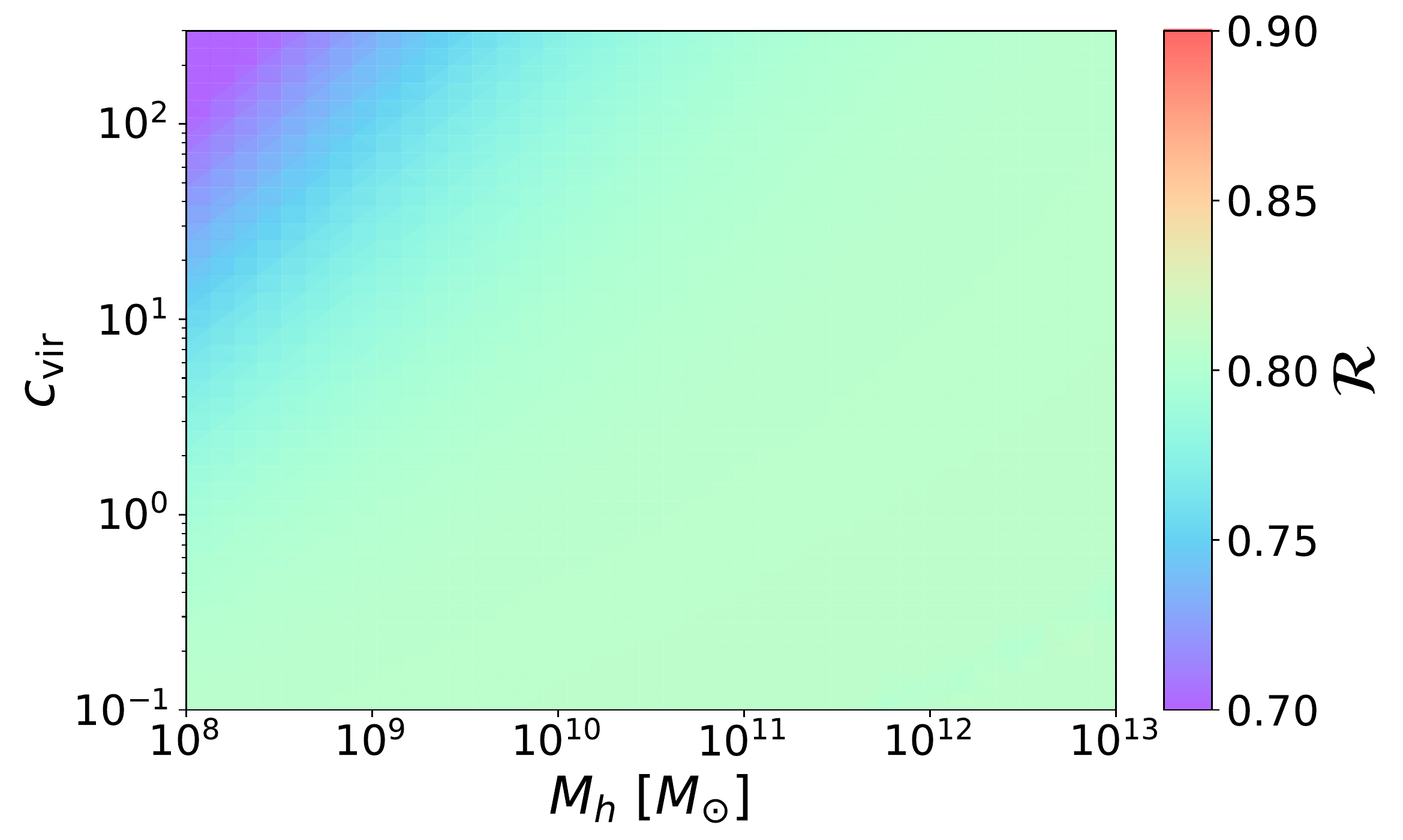}
\end{center}
\caption{{\it Left:} impact of soliton self-gravity on the core radius. Taking the backreaction of the soliton self-gravity into account, the estimated core radius is normalized by the one ignored the self-gravity, which we denote by $\RR$ (see Eq.~\ref{eq:ratio_core_radius}), and is plotted as a function of the dimensionless parameter $\alpha$. {\it Right}: same as left panel, but the result is plotted as function of the halo mass $M_{\rm h}$ and the concentration parameter $\cvir$, with the ratio $\RR$ depicted as color scales. Note that the mass of FDM is fixed to $m_\phi=3\times10^{-23}$\,eV and we follow the same parameter setup as adopted in Figs.~\ref{fig:core_halo_mass_concentration}--\ref{fig:core_halo_mass_CDM} (see Sec.~\ref{subsec:core-halo_mass-concentration_relation}). 
\label{fig:impact_core_radius}
}
\end{figure*}

Equation~(\ref{eq:ratio_core_radius}) implies that the self gravity of soliton always makes the core radius small. How it is small actually depends on the parameters, which are wholly encapsulated in the integral involving the ground-state eigenfunction and reduced potential. Note that these functions are solely characterized by the parameter $\alpha$. We thus first show the behavior of the ratio $\RR$ as a function of $\alpha$. The result is plotted in the left panel of Fig.~\ref{fig:impact_core_radius}. We then found that the ratio $\RR$ is nearly constant over a wide range of the parameter $\alpha$. As a result, the soliton self-gravity changes the core size by $20-30$\%. Here, for illustrative purpose, the plotted range of the parameter $\alpha$ was chosen to be sufficiently wide, $0.1\leq\alpha\leq10^5$. While this range partly includes the region where our analytical estimation of the soliton core size become inaccurate (i.e., $\alpha\lesssim1$), the resultant change in $\RR$ is fairly small, and no notable (sudden) transition appears. In the right panel, fixing the cosmology and FDM mass to those in the fiducial setup (see Sec.~\ref{subsec:core-halo_mass-concentration_relation}), the dependence on the halo mass and concentration parameter is shown for the case of $z=0$. The result depicted as color scales shows that the impact of soliton self gravity becomes gradually significant as decreasing the halo mass and increasing the halo concentration. As a result, the ratio $\RR$ reaches $\sim0.7$ as shown in the bluish upper triangular region. Note that at this region, the parameter $\alpha$ also becomes small and eventually below unity. However, the parameters of this region are rather extreme, and halos having such a high concentration parameter can never be realized in numerical simulations with a relevant cosmological setup.

Hence, we conclude that the soliton self-gravity can be non-negligible for the setup considered in Sec.~\ref{sec:result}, but its impact would not be large, suggesting that our analytical treatment ignoring the soliton self gravity is still valid, and can give a solid prediction for soliton core structure even from a quantitative viewpoint. Nevertheless, we note that our estimation given here relies on the perturbative treatment, and the contribution of the soliton potential is assumed to be small enough, compared to the halo potential. In this sense, the actual size of the impact may be quantitatively changed. For more accurate estimation, a nonperturbative approach by numerically solving the S-P equation is crucial, including self-consistently the soliton potential. This is left to our future study.  

\section{Conclusion}
\label{sec:conclusion}

In this paper, we have investigated analytically the structural properties of the soliton core, the stable dense core formed at the center of virialized halos in the fuzzy dark matter (FDM) model. Early numerical studies suggest that the size and mass of such a structure are tightly related to the host halo properties, given in a power-law form. On the other hand, more recent studies with high-resolution simulations reveal rich dynamical properties of the soliton core, also indicating that no tight relation between the soliton core structure and host halo mass exists, but instead there is a large scatter. To address these issues, we have presented an analytical description of the soliton core in a cosmological setup.

A crucial step to derive the soliton core properties is to solve the Schr\"odinger-Poisson (S-P) equation in an analytically tractable manner. In our treatment, we considered the gravitational potential of halo as a smooth background potential, and ignored the self-gravity of the soliton. With this setup, the problem to solve the S-P equation is reduced to a linear eigenvalue problem. Still, no exact solution is known, and to obtain analytically the eigenfunctions of soliton core, we have applied the Langer's method of uniform asymptotic approximation. We found that this method provides an excellent approximation which agrees well with the numerical solution of the linear eigenvalue problem, and thus gives an accurate description of the soliton eigenstates over a wide range of model parameters. Then, from the ground-state eigenfunction,  the key expression characterizing the soliton core radius was derived [Eqs.~(\ref{eq:x_core_ell_zero}) or (\ref{eq:r_core_halo_mass}) ], and combining the fitting formula for the core density $\rhoc$ estimated from simulation [Eq.~(\ref{eq:rhoc_Schive})], the expression for the soliton core mass was further obtained [Eq.~(\ref{eq:M_core_halo_mass})]. While these expressions apparently resemble those found in an early work \cite{Schive_etal2014b}, the newly derived formulas contain an additional factor, which depends not only on the cosmology and FDM mass but also on quantities characterizing the halo properties. Based on our new analytical formulas, a quantitative prediction of the soliton core-halo relations was made, and these were compared with results of numerical simulations of S-P equation.

Our important findings are summarized as follows:

\begin{itemize}
 \item The soliton core-halo relations, i.e., $\rc$-$M_{\rm h}$ and $M_{\rm core}$-$M_{\rm h}$ relations, depends sensitively not only on the cosmology and FDM mass but also on the halo concentration characterized by $\cvir$. In general, a lower concentration halo can host the soliton having a larger core radius and smaller core mass (see Fig.~\ref{fig:core_halo_mass_concentration}). Comparison of the predicted core-halo relations with simulation data suggests that the soliton core prefers halos with moderately small values of concentration parameter, $\cvir\sim\mathcal{O}(1-10)$. 

\item The predicted core-halo relations plotted as a function of halo concentration parameter (Fig.~\ref{fig:core_halo_mass_concentration}) shows that for each halo mass, there exist the allowable core size and core mass for the soliton states, respectively given as the upper and lower bounds. For fixed cosmological parameters and FDM mass, the theoretical upper limit on the core radius, $\rc^{\rm limit }$, and lower limit on the core mass, $M_{\rm c}^{\rm limit}$, are simply given as a function of halo mass, and are scaled as $\rc^{\rm limit }\propto M_{\rm h}^{-1/9}$ and $M_{\rm c}^{\rm limit}\propto M_{\rm h}^{1/9}$ [see Eqs.~(\ref{eq:r_core_upper_limit}) and (\ref{eq:M_core_lower_limit}) for more explicit expressions]. A general analysis in Appendix \ref{sec:soliton_core_cvir_limit} shows that these limits sensitively depend on the inner slope of halo density profile [see Eq.~(\ref{eq:rcore_limit_general})].

\item Incorporating explicitly the concentration-halo mass (C-M) relation into the predictions, the resultant core-halo relations generally deviate from a power-law form. In particular, the C-M relation monotonically decreasing with the halo mass, as typically seen in the CDM model, makes the core-halo relations flatter, apparently inconsistent with measured results in simulations (Fig.~\ref{fig:core_halo_mass_CDM}). On the other hand, the nonmonotonic C-M relation like the one indicated in the FDM model \cite{Dentler_etal2022} eventually exhibit a power-law feature at the small halo masses of $M_{\rm h}\sim10^8-10^9\,M_\odot$ (Fig.~\ref{fig:core_halo_mass_FDM}), leading to a close agreement with simulation results. 

\item Overall, the scatter around C-M relation produces a non-negligible amount of scatter in the soliton core-halo relations. This can explain the diversity advocated in Ref.~\cite{Jowett_etal2022}. An interesting trend seen in the simulation results may be that the scatter in the core-halo relations gets reduced as we go to small halo mass. This is rather consistent with predictions, in which a suppression of scatter is explained by the fact that the allowable size and mass of the soliton core are limited.  
\end{itemize}

We have also examined the validity of our treatment ignoring the soliton self-gravity. A perturbative estimation of its impact on the soliton core radius indicates that including the soliton self-gravity makes the effective potential of the S-P equation deep, leading to a shrinking of the core radius. The degree to which the core radius shrinks is characterized only by a single parameter, $\alpha$, and adopting the core density $\rhoc$ found in numerical simulations [Eq.~(\ref{eq:rhoc_Schive})], we found that the soliton self-gravity may change the results by $\sim20\%$, irrespective of the parameters we considered. Hence, the soliton self-gravity cannot be completely ignored, but its impact may not be large, and the predicted core-halo relations ignoring the self gravity would be still relevant, at least qualitatively, hence worthwhile for a further detailed comparison with new numerical simulations. Nevertheless, the validity of this estimation as well as a more proper description of soliton core have to be investigated, and we leave it to future work.

Finally, the analytical description presented in this paper would provide a useful and powerful route to characterize the dynamical properties of soliton core. As briefly mentioned in Secs.~\ref{sec:introduction} and \ref{subsec:spherical}, the soliton core seen in simulations is not strictly stable. Through the wave interference with granules distributed in a halo, the soliton core undergoes oscillations and random walks. These dynamical features can be indeed described as a superposition of the ground and excited states, including those having nonzero angular momenta \cite{Li_Hui_Yavetz2021, Luna_etal2022}. Although previous studies have numerically constructed such a dynamical state, the analytical construction of the eigenfunctions, described in Sec.~\ref{sec:Analytic_solutions_ell0} for $\ell=0$ and Appendix \ref{sec:Analytic_solutions_non_zero_ell} for $\ell\ne0$, would be advantageous to fast realize a dynamical soliton with less computational cost. Importantly, the present treatment provides a fairly general way to construct the eigenstates, and can be applied not only to the soliton in the NFW halo, but also to the one in a generalized halo profile expressed in an analytical form. Following proposals by Refs.~\cite{Lin_Schive_Wong_Chiueh2018,Yavets_Li_Hui2022}, one can even use our analytical treatment to generate the FDM halo structure consisting of many granules, each of which resembles the soliton eigenstates (see also Refs.~\cite{Dalal_Bovy_Hui_Li2021,Dandoy_Schwetz_Todarello2022}). Analytically describing the dynamical soliton states including halos is a very interesting and important subject and would offer a useful tool to observationally test the FDM model (e.g., Refs.~\cite{Kawai_etal2022,Dalal_Kravtsov2022} for related works). We will address it near future.

\begin{acknowledgments}
We are grateful to Kohei Hayashi and Hei Yin Jowett Chan for providing us their simulation data of soliton core structures and useful comments, Hiroki Kawai, Yusuke Manita, Takahiro Nishimichi, Masamune Oguri and Takuya Takahashi for discussions. We would like to also thank Mona Dentler and David J. Marsh for correcting the concentration-mass relation of fuzzy dark matter model used in this paper. This work was supported by Grant-in-Aid for JSPS Fellows No. 17J10553 (S. S.) and in part by MEXT/JSPS KAKENHI Grant Numbers JP20H05861 and JP21H01081 (A. T. ). A. T. acknowledges the support from JST 
AIP Acceleration Research Grant Number JP20317829. S. S. is supported by JSPS Overseas Research Fellowships. Numerical computation was carried out partly at the Yukawa Institute Computer Facility. 
\end{acknowledgments}

\appendix

\section{Halo density profile and mass concentration}
\label{app: NFW_profile}

In this Appendix, we summarize important characteristics of the dark matter halos used in the main text.

\subsection{Characteristics of halo density profile}
\label{subsec:NFW_halo}

The dark matter halo, a virialized self-gravitating dark matter clump,  is known to have a characteristic structure in density. In particular, a spherical mean of the density profile measured in cosmological $N$-body simulations is known to be quantitatively described by the Navarro-Frenk-White (NFW) profile \cite{Navarro_Frenk_White1996,Navarro_Frenk_White1997}. The comoving density of this profile is expressed as [Eq.~(\ref{eq:rho_NFW})]
\begin{align}
 \rho_{\rm NFW}(r)=\frac{\rhos}{(r/\rs)(1+r/\rs)^2},
\end{align}
where the radius $\rs$ describes the transition scale where the logarithmic slope of the density profile changes from $-3$ to $-1$ as we approach the halo center. The density $\rhos$ represents the characteristic density at $r=\rs$, and it is also expressed as follows\footnote{In the definition of $\rho_{\rm NFW}$ at Eq.~(\ref{eq:rho_NFW}), a factor of $1/a^3$ is factored out from the density profile. As a result, the dependence of $\rhos$ on the scale factor is somewhat different from the one in the literature.}:
\begin{align}
 \rhos=\frac{\Delta_{\rm vir}\,\rho_{\rm m,0}}{3}\,\frac{\cvir^3}{\ln(1+\cvir)-\cvir/(1+\cvir)}.
\label{eq:def_of_rhos}
\end{align}
Here, we introduced the concentration parameter $\cvir$ defined by $\cvir\equiv r_{\rm vir}/\rs$ with the radius $r_{\rm vir}$ being the virial radius. The virial radius is linked to the halo mass through 
\begin{align}
 r_{\rm vir}=\Bigl(\frac{3\,\mhalo}{4\pi\,\Delta_{\rm vir}\rho_{\rm m,0}}\Bigr)^{1/3}.
\label{eq:virial_radius}
\end{align}
The quantity $\Delta_{\rm vir}$ is the virial overdensity, which characterizes the mean overdensity of halos within the virial radius. While this is precisely described by the spherical collapse model \cite{Gunn_Gott1972,Gunn1977,Peebles:1980}, a simple and accurate fitting formula, which can be used in a flat universe with cosmological constant, is known \cite{Bryan_Norman1998} (see also Ref.~\cite{Nakamura_Suto1997}):  
\begin{align}
 \Delta_{\rm vir}(z)&=\frac{18\pi^2+82\,\{\Omega_{\rm m}(z)-1\}-39\{\Omega_{\rm m}(z)-1\}^2}{\Omega_{\rm m}(z)}.
\label{eq:virial_overdensity}
\end{align}
Here the quantity $\Omega_{\rm m}(z)$ is the matter density parameter at redshift $z$, given by  
\begin{align}
 \Omega_{\rm m}(z)&=\frac{(1+z)^3\,\Omega_{\rm m,0}}{(1+z)^3\Omega_{\rm m,0}+\Omega_{\Lambda,0}}
\label{eq:Omega_matter}
\end{align}
with the subscript $0$ indicating the quantities given at the present time. 

\subsection{Concentration-mass relations}
\label{subsec:C-M_relations}

For a quantitative characterization of the halo density profile, the key parameter is the concentration parameter $\cvir$, which characterizes the size and mass concentration of each halo. In the context of the CDM cosmology, there have been numerous works to quantify its cosmological and halo mass dependence based on the cosmological $N$-body simulations, and accurately calibrated models have been exploited. One of such representative models is given by Ref.~\cite{Bullock_etal2001}. The concentration parameter of this model, which we denote by $\cvir^{\rm B}$, is expressed as follows:
\begin{align}
 \cvir^{\rm B}(\mhalo,P_{\rm CDM})=A \frac{1+z_{\rm coll}(\mhalo,\,P_{\rm CDM})}{1+z},
\label{eq:cvir_Bullock}
\end{align}
with the constant $A$ chosen as $3.13$ \cite{Mead_etal2015}. The redshift $z_{\rm coll}$ characterizes the collapse time for halos with mass $\mhalo$ in the CDM cosmological model with the linear power spectrum $P_{\rm CDM}$. It is determined through 
\begin{align}
 \frac{D(z_{\rm coll})}{D(z)}\,\sigma(f_{\rm coll}\,\mhalo,\,P_{\rm CDM})=\delta_{\rm c},
\end{align}
where the function $D$ is the linear growth factor, and $\sigma$ is the root-mean-square amplitude of the linear density field smoothed with the top-hat filter function over the radius determined by the halo mass $\mhalo$, i.e., $\{3\,\mhalo/(4\pi\,\rho_{\rm m,0}\}^{1/3}$. The quantity $\delta_{\rm c}$ is the critical threshold for the linear density contrast, which we adopt the fitting form given by Ref.~\cite{Mead_etal2015}: $\delta_{\rm c}=1.59 + 0.0314 \ln \sigma_8(z)$ with $\sigma_8$ being the same rms fluctuation as we defined above, but smoothed at the specific radius of $8\,h^{-1}$\,Mpc. Here, the constant $f_{\rm coll}$ is set to $0.01$.

On the other hand, in the FDM cosmology, there is a little numerical work to calibrate the C-M relation based on the simulations of S-P equation. Still, there are several analytical works to model it in order to assess the small-scale problem as well as to constrain the FDM mass from weak lensing observations \cite{Marsh_Silk2014,Marsh_2016,Dentler_etal2022}. Here, we adopt the model presented by Ref.~\cite{Dentler_etal2022} and examine the core-halo relation based on our analytical treatment. The concentration parameter of this model is expressed as a correction to the model of Ref.~\cite{Bullock_etal2001} in the following form:
\begin{align}
 \cvir^{\rm F}(\mhalo) = \cvir^{\rm B}(\mhalo,P_{\rm FDM})\Delta_{\rm c}^{\rm FDM},
\end{align}
where the concentration parameter $\cvir^{\rm B}$ is obtained from Eq.~(\ref{eq:cvir_Bullock}) with the linear power spectrum $P_{\rm FDM}$ given in the FDM cosmological model [see e.g., Eqs.~(8) with (9) in Ref.~\cite{Hu_Barkana_Gruzinov2000}]. The correction term $\Delta_{\rm c}^{\rm FDM}$ is defined by 
\begin{align}
& \Delta_{\rm c}^{\rm FDM}=\Bigl\{1+\frac{M_0^{\rm (c)}}{\mhalo}\Bigr\}^{-\gamma_0}\,\Bigl\{1+\gamma_1\,\frac{f_{\rm coll}\,M_0^{\rm (c)}}{\mhalo}\Bigr\}^{-\gamma_2},
\end{align}
with the coefficient $\gamma_1$ defined by $\gamma_1=15$ and indices $\gamma_0$ and $\gamma_2$, respectively, given by $\gamma_0=d \ln\cvir^{\rm B}/d\ln \mhalo|_{4\,M_0^{\rm (c)}}$ and $\gamma_2=0.3$. Here, the characteristic mass $M_0^{\rm (c)}$ is related to the scaling mass $M_0^{\rm (n)}$ through $M_0^{\rm(c)}=M_0^{\rm(n)}/f_{\rm coll}$, and $M_0^{\rm (n)}$ is defined by
\begin{align}
M_0^{\rm (n)}=1.6\times10^{10}\,\,M_\odot\,\Bigl(\frac{m_\phi}{10^{-22}\,{\rm eV}}\Bigr)^{-4/3}.
\end{align}

\section{Analytical solutions $\ell\ne 0$}
\label{sec:Analytic_solutions_non_zero_ell}

In this Appendix, extending the treatment in Sec.~\ref{subsec:Langer_approx}, we construct analytically the approximate solutions of the S-P equation given at Eq.~(\ref{eq:stationary-SP_eq2}) in the case of $\ell\ne0$.

\subsection{Constructing analytical Eigenstates}
\label{subsec:eigenstates_ell_non-zero}

Similarly to the $\ell=0$ case in Sec.~\ref{subsec:Langer_approx}, we rewrite Eq.~(\ref{eq:stationary-SP_eq2}) with the normal form, keeping the angular momentum nonzero. Again introducing the new function $\tilde{u}_{n\ell}\equiv x \,u_{n\ell}$, the equation for the radial wave function is given in the following form:
\begin{align}
& \frac{d^2\tilde{u}_{n\ell}(x)}{dx^2}+\alpha\,g_\ell(x)\tilde{u}_{n\ell}(x)=0\,;
\nonumber
\\
&\qquad \quad
g_\ell(x)\equiv -\frac{\ell(\ell+1)}{\alpha\,x^2}+\frac{\log(1+x)}{x}+\frac{\mathcal{E}}{\alpha}.
\label{eq:SP_eq3_ell}
\end{align}
Imposing the same boundary conditions as given at Eq.~(\ref{eq:boundary_condition}), we construct analytically the eigenfunction and eigenvalues of this system. Following the same steps as we examined in the $\ell=0$ case, we apply the Liouville-Green transformation $(x,\tilde{u}_{n\ell})\to(z,v_{n\ell})$ in Eq.~(\ref{eq:Liouville-Green_transformation}). Equation~(\ref{eq:SP_eq3_ell}) is then rewritten with 
\begin{align}
 \frac{d^2v_{n\ell}(z)}{dz^2}+\Bigl[\alpha\frac{g_\ell(x)}{\{p'(x)\}^2}+\delta\Bigr]v_{n\ell}(z)=0,
\label{eq:SP_eq4_ell}
\end{align}
where the quantity $\delta$ is defined by Eq.~(\ref{eq:def_delta}).

Comparing the above equation with Eq.~(\ref{eq:SP_eq4_ell0}) in the $\ell=0$ case, an important difference appears in the function $g_\ell$, which has now the two turning points. That is, the function becomes zero at $x=x_1$ and $x_2$, and is positive and negative, respectively, at $x_1\leq x \leq x_2$ and $x<x_1$ or $x>x_2$.  In order to obtain a uniformly valid approximation, the function $z=p(x)$ has to be chosen in such a way that it behaves like the function $g_\ell$, having two zero-crossing points. Here, we specify the functional form of $z$ in the following way \cite{Langer1959,Olver1975}:
\begin{align}
 \frac{g_\ell(x)}{\{p'(x)\}^2}&=\beta^2-z^2
\nonumber
\\
&=\beta^2-\{p(x)\}^2, 
\label{eq:choice_p2}
\end{align}
where the quantity $\beta$ is a constant which will be given later. 
Equation~(\ref{eq:choice_p2}) is the first-order differential equation for the function $p$, which is analytically integrated to give the following relations:
\begin{align}
\int_{x_1}^x\sqrt{g_\ell(x')}\,dx'&=\int_{-\beta}^p \sqrt{\beta^2-\tau^2}\,d\tau\nonumber
\\ 
&=\frac{1}{2}\beta^2\,\arccos\Bigl(-\frac{p}{\beta}\Bigr)+\frac{p}{2}\,\sqrt{\beta^2-p^2}
\label{eq:integ_x1_x_x2}
\end{align}
for $x_1<x<x_2$, 
\begin{align}
\int_x^{x_1}\sqrt{-g_\ell(x')}\,dx'&=\int_p^{-\beta} \sqrt{\tau^2-\beta^2}\,d\tau\nonumber
\\ 
&=-\frac{1}{2}\Bigl\{\beta^2\,\cosh^{-1}\Bigl(-\frac{p}{\beta}\Bigr)+p\,\sqrt{p^2-\beta^2}\Bigr\}
\label{eq:integ_x_x1}
\end{align}
for $x\leq x_1$, and
\begin{align}
\int_{x_2}^x\sqrt{-g_\ell(x')}\,dx'&=\int_\beta^p \sqrt{\tau^2-\beta^2}\,d\tau\nonumber
\\ 
&=-\frac{1}{2}\Bigl\{\beta^2\,\cosh^{-1}\Bigl(\frac{p}{\beta}\Bigr)-p\,\sqrt{p^2-\beta^2}\Bigr\}
\label{eq:integ_x2_x}
\end{align}
for $x\geq x_2$.

From Eqs.~(\ref{eq:integ_x1_x_x2})-(\ref{eq:integ_x2_x}) given above, one can construct the function $p$ expressed in terms of the variable $x$ as follows. Provided the functional form of $g_\ell$, the integrals at the left-hand side of these equations are numerically performed, and the results are given as a function of $x$. On the other hand, the functions at the right-hand side of Eqs.~(\ref{eq:integ_x1_x_x2})-(\ref{eq:integ_x2_x}) are evaluated as a function of $p$. That is, Eqs.~(\ref{eq:integ_x1_x_x2})-(\ref{eq:integ_x2_x}) are symbolically written in the form as $A(x)=B(p)$. Hence, solving numerically this implicit equation for $p$, i.e., $p=B^{-1}(A(x))$, the functional form of $p(x)$ is determined. In the above, the constant $\beta$ is arbitrary, but we shall specify it in such a way that with the Liouville-Green transformation, the turning points $x_1$ and $x_2$ are mapped to $p=-\beta$ and $p=\beta$, respectively. From Eq.~(\ref{eq:integ_x1_x_x2}), this is equivalent to imposing the following condition:
\begin{align}
\beta^2=\frac{2}{\pi}\int_{x_1}^{x_2}\sqrt{g_\ell(x')}\,dx'. 
\label{eq:beta_func_g_ell}
\end{align}

With the specific choice at Eq.~(\ref{eq:choice_p2}), Eq.~(\ref{eq:SP_eq4_ell}) is now recast as
\begin{align}
 \frac{d^2v_{n\ell}(z)}{dz^2}+\alpha\bigl(\beta^2-z^2\bigr)v_{n\ell}(z)=-\delta(z)v_{n\ell}(z).
\end{align}
Ignoring the term at right-hand side, the general solution of this equation is described by the parabolic cylinder functions $U$ and $\overline{U}$ \cite{Olver1975,Langer1959}\footnote{The function $U$ given here is exactly the same one as defined in Ref.~\cite{Abramowitz_Stegun1965}. Also, it is related to the Weber's parabolic cylinder function $D_\lambda$ through $D_\lambda(x)=U(-\lambda-1/2,x)$. On the other hand, the function $\overline{U}$ slightly differ from the $V$ in Ref.~\cite{Abramowitz_Stegun1965} by the factor of $\Gamma(1/2-b)$, i.e., $U(b,x)=\Gamma(1/2-b)\,V(a,x)$.}: 
\begin{align}
 v_{n\ell}(z) &=\tilde{c}_1 \,U\Bigl(-\frac{1}{2}\alpha^{1/2}\beta^2,\,\sqrt{2\alpha^{1/2}}\,z\Bigr)
\nonumber
\\
&\qquad+\tilde{c}_2 \,\overline{U}\Bigl(-\frac{1}{2}\alpha^{1/2}\beta^2,\,\sqrt{2\alpha^{1/2}}\,z\Bigr)
\label{eq:general_solution_v_non_zero_ell}
\end{align}
or
\begin{align}
 \tilde{u}_{n\ell}(x) &=\Biggl[\frac{|\beta^2-\{z(x)\}^2|}{|g_\ell(x)|}\Biggr]^{1/4}
\nonumber
\\
&\times
\Biggl\{c_1 \,U\Bigl(-\frac{1}{2}\alpha^{1/2}\beta^2,\,\sqrt{2\alpha^{1/2}}\,z(x)\Bigr)
\nonumber
\\
&\quad\quad 
+c_2 \,\overline{U}\Bigl(-\frac{1}{2}\alpha^{1/2}\beta^2,\,\sqrt{2\alpha^{1/2}}\,z(x)\Bigr)\Biggr\},
\label{eq:general_solution_utilde_non_zero_ell}
\end{align}
where the coefficients $\tilde{c}_i$ and $c_i$ are the integration constants.

Note that the parabolic cylinder functions have the following asymptotic behaviors:
\begin{align}
 U(b,\zeta)&\stackrel{\zeta\to\infty}{\sim} \frac{1}{\zeta^{b+1/2}}\,e^{-\zeta^2/4},
\nonumber
\\
 \overline{U}(b,\zeta)&\stackrel{\zeta\to\infty}{\sim} \frac{2^{1/2}\,\zeta^{b-1/2}}{\pi^{1/2}}\,\Gamma\Bigl(\frac{1}{2}-b\Bigr)\,e^{\zeta^2/4},
\label{eq:pbc_func_asymptotic1}
\end{align}
and
\begin{align}
 U(b,\zeta)\stackrel{\zeta\to-\infty}{\sim} \frac{(2\pi)^{1/2}(-\zeta)^{b-1/2}}{\Gamma(1/2+b)}\,e^{\zeta^2/4}.
\label{eq:pbc_func_asymptotic2}
\end{align}
Let us recall from Eqs.~(\ref{eq:integ_x1_x_x2})--(\ref{eq:beta_func_g_ell}) that the function $p(x)$ or $z(x)$ is a monotonically increasing function of $x$, and varies from negative to positive. Thus, the quantity $p$ approaches a negative constant value at $x=0$ and goes to infinity in the limit of $x\to\infty$. From Eq.~(\ref{eq:pbc_func_asymptotic1}),  this indicates that the second term in Eqs.~(\ref{eq:general_solution_v_non_zero_ell}) or (\ref{eq:general_solution_utilde_non_zero_ell}) diverge in the limit of $x\to+\infty$. Hence, to fulfill the boundary condition $\tilde{u}(\infty)=0$,  the coefficients $\overline{c}_2$ and $c_2$ must be zero.

On the other hand, another boundary condition $\tilde{u}(0)=0$ yields 
\begin{align}
 U\Bigl(-\frac{1}{2}\alpha^{1/2}\beta^2,\,\sqrt{2\alpha^{1/2}}\,z(0)\Bigr)=0, 
\label{eq:eigenvalu_Parabolic_cylinder}
\end{align}
where we used the fact that the functions $g_{\ell}(x)$ and $|\beta^2-\{z(x)\}^2|$ are nonzero at $x=0$. Equation~(\ref{eq:eigenvalu_Parabolic_cylinder}) is the transcendental equation for the eigenvalue $\varepsilon/\alpha$. That is, given the parameters $\alpha$ and $\ell$, the turning points $x_{1,2}$ are determined, and the constant $\beta$ and quantity $z(0)$ are computed, with both of them given as a function of $\mathcal{E}/\alpha$. Then, substituting them into Eq.~(\ref{eq:eigenvalu_Parabolic_cylinder}), solving it yields a discrete set of eigenvalues $\mathcal{E}/\alpha$. 

Once the eigenvalues are determined from Eq.~(\ref{eq:eigenvalu_Parabolic_cylinder}), the functional form of the eigenfunctions for $\ell\ne0$ are specified in an unambiguous way except for the normalization factor. To sum up, the (unnormalized) eigenfunctions for $\ell\ne0$ that are constructed analytically are expressed as 
\begin{widetext}
\begin{align}
 \tilde{u}_{n\ell}(x)&=x\,u_{n\ell}(x)
=\Bigl[\frac{|\beta^2-\{z(x)\}^2|}{|g_\ell(x)|}\Bigr]^{1/4}\,U\Bigl(-\frac{1}{2}\alpha^{1/2}\beta^2,\,\sqrt{2\alpha^{1/2}}\,z(x)\Bigr),\,\,\,
& \beta^2=\frac{2}{\pi}\int_{x_1}^{x_2}\sqrt{g_\ell(x')}\,dx'
\label{eq:approx_eigenfunction_non_zero_ell}
\end{align}
\end{widetext}
with the function $g_\ell$ given at Eq.~(\ref{eq:SP_eq3_ell}). The function $p=z(x)$ is obtained by Eqs.~(\ref{eq:integ_x1_x_x2})--(\ref{eq:integ_x2_x}).

\begin{table*}[tb]
 \caption{Lowest five eigenvalues normalized by $\alpha$ in the cases of the nonzero angular momenta $\ell=2$, $5$ and $10$, setting the parameter $\alpha$ to $\alpha=10^3$ (see Fig.~\ref{fig:eigen_func_non-zero_ell}). Analytical results obtained from Eq.~(\ref{eq:eigenvalu_Parabolic_cylinder}) are summarized. For comparison, parentheses show the results by solving numerically the matrix eigenvalue problem, adopting the boundaries $x_1=0$ and $x_n=5$ with the number of grids $n=10^4$.
\label{tab:eigenvalues_non-zero_ell}}
\begin{ruledtabular}
\begin{tabular}{lccc}
$\mathcal{E}_{n\ell}/\alpha$ & $\ell=2$ & $\ell=5$ & $\ell=10$
\\
\hline
$n=1$  & -0.77454 (-0.77278) &-0.67810 (-0.67756) & -0.57018 (-0.56999)
\\
$n=2$  & -0.71757 (-0.71620) &-0.63909 (-0.63863) & -0.54483 (-0.54466)
\\
$n=3$  & -0.67142 (-0.67029) &-0.60531 (-0.60491) & -0.52189 (-0.52174)
\\
$n=4$  & -0.63259 (-0.63163) &-0.57561 (-0.57526) & -0.50100 (-0.50087)
\\
$n=5$  & -0.59911 (-0.59828) &-0.54916 (-0.54885) & -0.48187 (-0.48175)
\\
\hline
\end{tabular}
\end{ruledtabular}
\end{table*}

\begin{figure*}[tb]
 \includegraphics[width=18cm,angle=0]{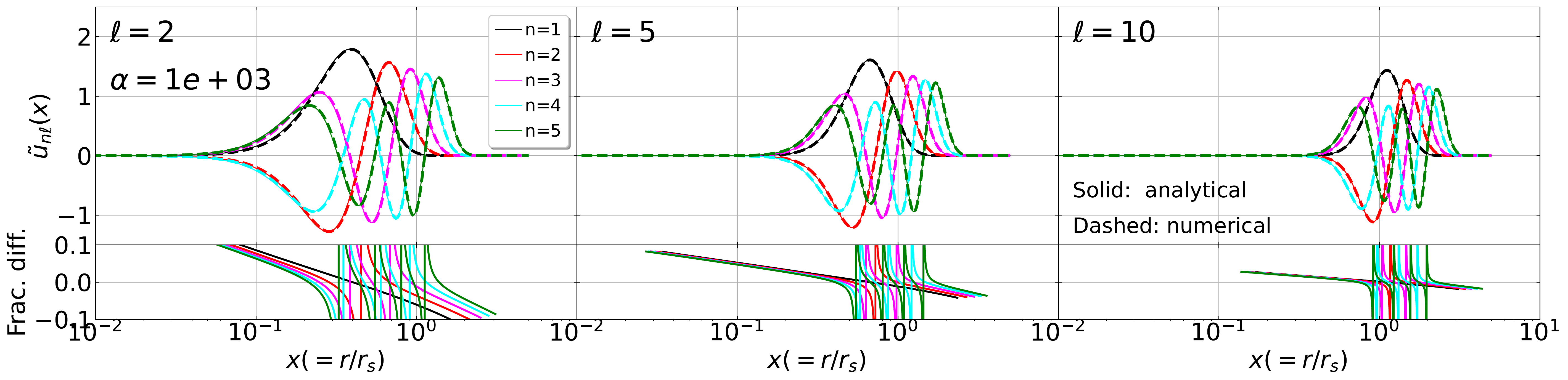}
\caption{Eigenfunctions for the radial wave function having nonzero angular momentum, $\tilde{u}_{n\ell}(x)$. From left to right, fixing the parameter $\alpha$ to $10^3$, eigenfunctions for the lowest five eigenstates are shown for $\ell=2$, $5$, and $10$, respectively. In the upper panels, thin solid lines are the eigenfunctions constructed analytically, while the thick dashed lines are the results obtained by solving the matrix eigenvalue problem numerically. Both of them are properly normalized. Bottom panels show the fractional difference between analytical and numerical results, defined by $(\tilde{u}_{n\ell,{\rm analytical}}-\tilde{u}_{n\ell,{\rm numerical}})/\tilde{u}_{n\ell,{\rm numerical}}$ (see Fig.~\ref{fig:eigen_func_alpha10_100_1000}). 
Note that to avoid the division by zero, we stop plotting the results when the amplitude of the wave functions becomes smaller than $10^{-6}$ in the inner and outer regions.
\label{fig:eigen_func_non-zero_ell}
}
 \includegraphics[width=18cm,angle=0]{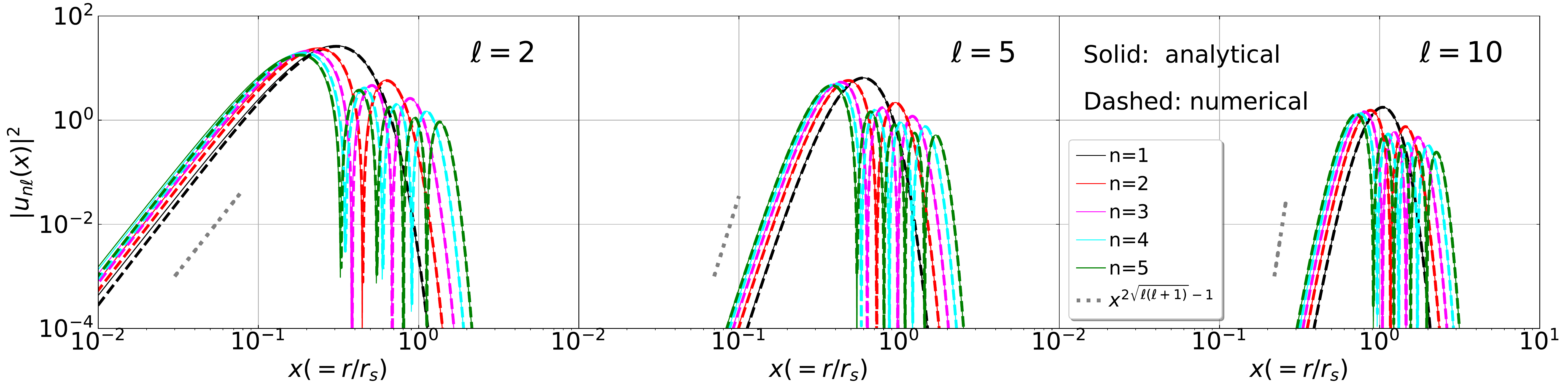}
\caption{
Same as Fig.~\ref{fig:eigen_func_non-zero_ell}, but we plot the square of the normalized wave function, $u_{n\ell}(x)=x\,\tilde{u}_{n\ell}(x)$, i.e., $|u_{n\ell}(x)|^2$. Line and color types shown hare are the same as those used in Fig.~\ref{fig:eigen_func_non-zero_ell}. In each panel, the gray dashed lines are the asymptotic slope of the wave function at $x\lesssim1$ predicted by the WKB approximation [see Eq.~(\ref{eq:inner_slope_non_zero_ell})].
\label{fig:eigen_func_non-zero_ell_log}
}
\end{figure*}

\subsection{Comparison with numerical solutions}
\label{subsec:comparison_non-zero_ell}

Let us compare the analytical eigenstates for $\ell\ne0$ to the numerical results obtained by solving the matrix eigenvalue problem. For illustration, we below focus on the specific angular momenta of $\ell=2$, $5$, and $10$. 

Table \ref{tab:eigenvalues_non-zero_ell} presents the lowest five eigenvalues for non-zero angular momenta, fixing the dimensionless parameter $\alpha$ to $\alpha=10^3$. In Figs.~\ref{fig:eigen_func_non-zero_ell} and \ref{fig:eigen_func_non-zero_ell_log}, the corresponding (normalized) eigenfunctions $\tilde{u}_{n\ell}$ and the square of eigenfunctions, $|u_{n\ell}|^2$, are, respectively, plotted as a function of $x=r/r_s$. Line types and colors are the same as in Fig.~\ref{fig:eigen_func_alpha10_100_1000}. Note that in obtaining the numerical results, we solve Eq.~(\ref{eq:matrix_eigenvalue_problem}), with the potential $V(x)$ being replaced with $V(x)=-\alpha\,\log(1+x)/x +\ell(\ell+1)/x^2$. 

Overall, we see an excellent agreement between analytical and numerical results, similarly to the $\ell=0$ case. Although a closer look at the fractional difference shown in the bottom panels of Fig.~\ref{fig:eigen_func_non-zero_ell} reveals that the eigenfunctions exhibit discrepancies, these become non-negligible only when the eigenfunction becomes close to zero, and hence their net impact is considered to be small. Hence, we conclude that the analytically constructed eigenstates are accurate enough even for the nonzero angular momenta, and would be useful to characterize quantitatively the dynamical properties of the soliton core.

Finally, one notable behavior in the square of the eigenfunction, corresponding to the soliton density profile, is the characteristic peak structure shown in Fig.~\ref{fig:eigen_func_non-zero_ell_log}. As increasing $\ell$, the density peak generically appears at a larger radius, with its width narrower. In particular, the inner structure exhibits a power-law behavior, which we find to be scaled as $|u_{n\ell}|^2\propto x^{2\sqrt{\ell(\ell+1)}-1}$ at $x\ll1$. This is regardless of whether it is in the ground or excited state (see gray dashed lines). In Appendix \ref{subsec:WKB_approx}, we derive this power-law form explicitly based on the WKB approximation.

\subsection{Asymptotic behaviors near origin}
\label{subsec:WKB_approx}

In this Appendix, we discuss the asymptotic behavior of the eigenfunction $\tilde{u}_{n\ell}$ near the halo center, and derive analytically the power-law form of the density profile as shown in  Fig.~\ref{fig:eigen_func_non-zero_ell_log}.  In doing so, rather than using Eq.~(\ref{eq:approx_eigenfunction_non_zero_ell}), we apply the WKB approximation and obtain an analytical expression for the eigenfunction valid near the origin $(x\approx0)$, where the function $g_\ell(x)$ is supposed to be negative. The (unnormalized) WKB solution is then given in the form as
\begin{align}
& \tilde{u}_{n\ell}^{\rm WKB}(x)\simeq \frac{1}{\{-g_\ell(x)\}^{1/4}}\,
\nonumber
\\
&\quad\times
\Biggl[
d_1\,\exp\Bigl\{\alpha^{1/2}\int_x^{x_1}  \sqrt{-g_\ell(x')} dx' \Bigr\}
\nonumber
\\
&\quad +
d_2\,\exp\Bigl\{-\alpha^{1/2}\int_x^{x_1}  \sqrt{-g_\ell(x')} dx' \Bigr\}
\Biggr]
\label{eq:WKB_around_x=0}
\end{align}
with the upper bound of the integral, $x_1$, being the first turning point, $g_\ell(x_1)=0$. In the above, the coefficients $d_1$ and $d_2$ are not arbitrary, but are related with each other through the inner boundary condition, $\tilde{u}_{n\ell}(0)=0$. This relation is rewritten in the following form:
\begin{align}
\frac{d_1}{d_2}=-\exp\Bigl[-2\alpha^{1/2}\int^{x_1}_0\,\sqrt{-g_\ell(x')}\,dx'\Bigr].
\label{eq:relation_d1_d2}
\end{align}
Recalling that near the origin, the function $g_\ell$ is approximated as $g_\ell\simeq\ell(\ell+1)/x^2/\alpha$, the integral in the exponent above, $\int^{x_1}_0\,\sqrt{-g_\ell(x')}\,dx'$, is shown to exhibit a logarithmic divergence. To be precise, we may replace the lower bound of this integral with a nonzero small and positive value of $\eta$, and perform the integration. Assuming $x_1\ll1$, this leads to $\sqrt{\ell(\ell+1)/\alpha}\log(x_1/\eta)$. Taking the limit $\eta\to0$, the integral thus diverges positively. Note that the logarithmic divergence generically appears regardless of whether $x_1$ is small or not. In other words, the right-hand side of Eq.~(\ref{eq:relation_d1_d2}) becomes vanishing, and hence the coefficient $d_1$ must be zero. Equation~(\ref{eq:WKB_around_x=0}) is recast as
\begin{align}
 \tilde{u}_{n\ell}^{\rm WKB}(x)= \frac{d_2}{\{-g_\ell(x)\}^{1/4}}\,
\,\exp\Bigl\{-\alpha^{1/2}\int_x^{x_1}  \sqrt{-g_\ell(x')} dx' \Bigr\}.
\label{eq:WKB_around_x=0_2}
\end{align}

With the wave function in Eq.~(\ref{eq:WKB_around_x=0_2}), we derive an asymptotic power-law form at $x\ll1$. To obtain an expression without assuming a specific value of $x_1$,  we consider a sufficiently small but still finite value of $x_0\,(\ll x_1)$, below which the function $g_\ell(x)$ is approximately described by $g_\ell(x)\simeq \ell(\ell+1)/x^2/\alpha$. Focusing on the region of $x\lesssim x_0$, the integral in the exponent of Eq.~(\ref{eq:WKB_around_x=0_2}) is evaluated to give
\begin{align}
&\int_x^{x_1}\sqrt{ -g_\ell(x')}\,dx' 
\nonumber
\\
&\quad=\int_{x_0}^{x_1}\sqrt{ -g_\ell(x')}\,dx'
+\frac{\sqrt{\ell(\ell+1)}}{\alpha^{1/2}}\,\log\Bigl(\frac{x_0}{x}\Bigr).
\end{align}
Plugging this into Eq.~(\ref{eq:WKB_around_x=0_2}), 
the WKB solution near the origin $(x\lesssim x_0)$ is rewritten in the following form:
\begin{widetext}
\begin{align}
\tilde{u}_{n\ell}^{\rm WKB}(x)&\simeq 
d_2\,\frac{\alpha^{1/4}\,}{\{\ell(\ell+1)\}^{1/4}}\,x^{1/2}\,\Bigl(\frac{x}{x_0}\Bigr)^{\sqrt{\ell(\ell+1)}}\,
\exp\Bigl\{-\alpha^{1/2}\int_{x_0}^{x_1}  \sqrt{-g_\ell(x')} dx'\Bigr\}.
\label{eq:inner_WKB_non_zero_ell}
\end{align}
\end{widetext}
This shows that the density profile for the soliton state, proportional to $|u_{n\ell}|^2$, exhibits a power-law behavior near the origin and it is proportional to 
\begin{align}
 |u_{n\ell}^{\rm WKB}(x)|^2=\Bigl|\frac{\tilde{u}_{n\ell}^{\rm WKB}(x)}{x}\Bigr|^2\propto x^{2\sqrt{\ell(\ell+1)}-1},\quad (x\ll x_0)
\label{eq:inner_slope_non_zero_ell}
\end{align}
Since the above expression has no explicit dependence on $n$, this scaling generally holds for both ground and excited states. Note cautiously that in deriving Eq.~(\ref{eq:inner_slope_non_zero_ell}), we assumed nonzero $\ell$ and used the asymptotic behavior of $g_\ell$. In this respect, Eqs.~(\ref{eq:inner_WKB_non_zero_ell}) and (\ref{eq:inner_slope_non_zero_ell}) are valid only for $\ell\ne0$ and cannot be applied to the $\ell=0$ case.

\section{Soliton core size in the limit of $\cvir\to0$}
\label{sec:soliton_core_cvir_limit}

In this Appendix, we discuss the dependence of the soliton core size on the halo density profile. To be specific,  we investigate how the upper bound on the core radius, which is obtained by taking the limit $\cvir\to0$, varies with the inner slope of density profile. 

Consider a class of spherically symmetric halos having a characteristic scale $\rs$, below which the radial profile asymptotically follows a power-law form. One can write such a profile formally as
\begin{align}
 \rho(x)=\rhos\,D(x), \,\,\,x=r/\rs
\label{eq:profile_general}
\end{align}
with the dimensionless function $D$ asymptotically behaving as
\begin{align}
 D(x)\stackrel{x\ll1}{\simeq} x^{-s}.
\end{align}
Note that Eq.~(\ref{eq:profile_general}) is the comoving density, and
a factor of $1/a^3$ is factorized in the same way as defined in Eq.~(\ref{eq:rho_NFW}). Thus, for the NFW profile, $\rhos$ is given by Eq.~(\ref{eq:def_of_rhos}) and $D$ becomes $D(x)=1/x/(1+x)^2$. The mass and potential profiles of this model are, respectively, given by
\begin{align}
 M(x) &= 4\pi\,\rhos\,\rs^3\int_0^x dy\, y^2 D(y),
\label{eq:mass_general}
\\
 \Psi(x) &= -\frac{4\pi\,G\,\rhos\,\rs^2}{a}\int_x^\infty \frac{dy}{y^2}\int_0^y dw\, w^2\,D(w)
\nonumber
\\
&\equiv -\frac{4\pi\,G\,\rhos\,\rs^2}{a}\,\tilde{\Psi}(x).
\label{eq:potential_general}
\end{align}
Below, we shall consider the case that both the mass and potential are finite at the center. This implies the inner slope $s$ has to be less than $2$. In such a case, the dimensionless potential $\tilde{\Psi}$ at $x\ll1$ is generally written in the following form: 
\begin{align}
 \tilde{\Psi}(x)\simeq \tilde{\Psi}(0) - \frac{x^{2-s}}{(3-s)(2-s)},\quad(s<2). 
\label{eq:tilde_Phi_near_origin}
\end{align}
With the halo mass defined by $M_{\rm h}=(4\pi/3)\rho_{\rm m,0}\Delta_{\rm vir} r_{\rm vir}^3$ at $r=r_{\rm vir}$, the characteristic density $\rhos$ and the scale radius $\rs$ are expressed in terms of the concentration parameter $\cvir=r_{\rm vir}/\rs$ as follows:
\begin{align}
 \rhos &=\frac{\rho_{\rm m,0}\Delta_{\rm vir}}{3}\,\cvir^3\Bigl\{\int_0^{c_{\rm vir}} y^2D(y)dy\Bigr\}^{-1},
\nonumber
\\
&\simeq \frac{\rho_{\rm m,0}\Delta_{\rm vir}}{3} (3-s)\,\cvir^{s},
\label{eq:rhos_cvir_general}
\\
  \rs &=\cvir^{-1}\,\Bigl(\frac{3}{4\pi}\frac{M_{\rm h}}{\rho_{\rm m,0}\Delta_{\rm vir}}\Bigr)^{1/3},
\label{eq:rs_cvir_general}
\end{align}
where the last line of Eq.~(\ref{eq:rhos_cvir_general}) is valid for $\cvir\ll1$.

Provided basic quantities to characterize the halos, we estimate the soliton core size in the limit of $\cvir\to0$. As we have discussed in Sec.~\ref{sec:Analytic_solutions_ell0}, the S-P equation ignoring the soliton self-gravity is reduced to a linear eigenvalue problem, and focusing on $\ell=0$, the governing equation is recast in the form similar to Eq.~(\ref{eq:SP_eq3_ell0}), but the halo potential is replaced with Eq.~(\ref{eq:potential_general}). Nevertheless, this modification does not affect the monotonicity of the halo potential, and we can follow the same steps as in the case of the NFW profile to analytically construct the eigenstates. At the end, we obtain the same analytical expression as given at Eq.~(\ref{eq:eigen_func_analytical_ell0}), from which the expression of soliton core size can be derived based on the asymptotic properties of the Airy function (see Sec.~\ref{subsec:core-structure}). We have\footnote{In deriving Eq.~(\ref{eq:x_core_general}), we consider only the leading-order term proportional to $\alpha$, and ignore other contributions. }: 
\begin{align}
 \xc=p\sqrt{\frac{6}{\alpha\,g_0(0)}} \,;\quad p=0.65,
\label{eq:x_core_general}
\end{align}
with the function $g_0(x)$ given by
\begin{align}
 g_0(x)=\tilde{\Psi}(x) + \mathcal{E}/\alpha.
\label{eq:func_g0_general}
\end{align}
The above expression is regarded as a generalized formula for the core radius. From this, the comoving core radius $\rc$ is written as 
\begin{align}
 \rc&=\xc\,\rs 
\nonumber
\\
&= p\sqrt{\frac{6}{\bigl\{\tilde{\Psi}(0)+\mathcal{E}/\alpha\bigr\}}}\,\frac{\rs}{\alpha^{1/2}}
\label{eq:r_core_general}
\end{align}

In the above, the parameter $\alpha$ is defined similarly as in the case of the NFW profile, i.e., $\alpha=8\pi\,G\,\mass^2\,\rhos\,\rs^4\,a$ [see Eq.~(\ref{eq:parameters})]. Importantly, however, its dependence on the concentration parameter differs generally from the one given in Eq.~(\ref{eq:alpha_value}). To explicitly see this, we use Eqs.~(\ref{eq:rhos_cvir_general}) and (\ref{eq:rs_cvir_general}) to rewrite the expression of $\alpha$ with 
\begin{align}
 \alpha 
&= 8\pi\,G\,m_\phi^2\,a \frac{\rho_{\rm m,0}\Delta_{\rm vir}}{3}
\Bigl( \frac{3}{4\pi}\frac{M_{\rm h}}{\rho_{\rm m,0}\Delta_{\rm vir}}\Bigr)^{4/3}
\nonumber
\\
&\quad\qquad\times\Bigl\{ \cvir \int_0^{c_{\rm vir}}y^2\,D(y)dy \Bigr\}^{-1}
\nonumber
\end{align}
Thus, taking the limit $\cvir\ll1$, we obtain
\begin{align}
\alpha \simeq 
\Bigl(\frac{6}{\pi}\Bigr)^{1/3}G\,m_\phi^2\, a \Bigl(\frac{M_{\rm h}^4}{\rho_{\rm m,0}\Delta_{\rm vir}}\Bigr)^{1/3} (3-s)\,\cvir^{s-4}.
\label{eq:alpha_general}
\end{align}

\begin{figure}[tb]
 \includegraphics[width=7.5cm,angle=0]{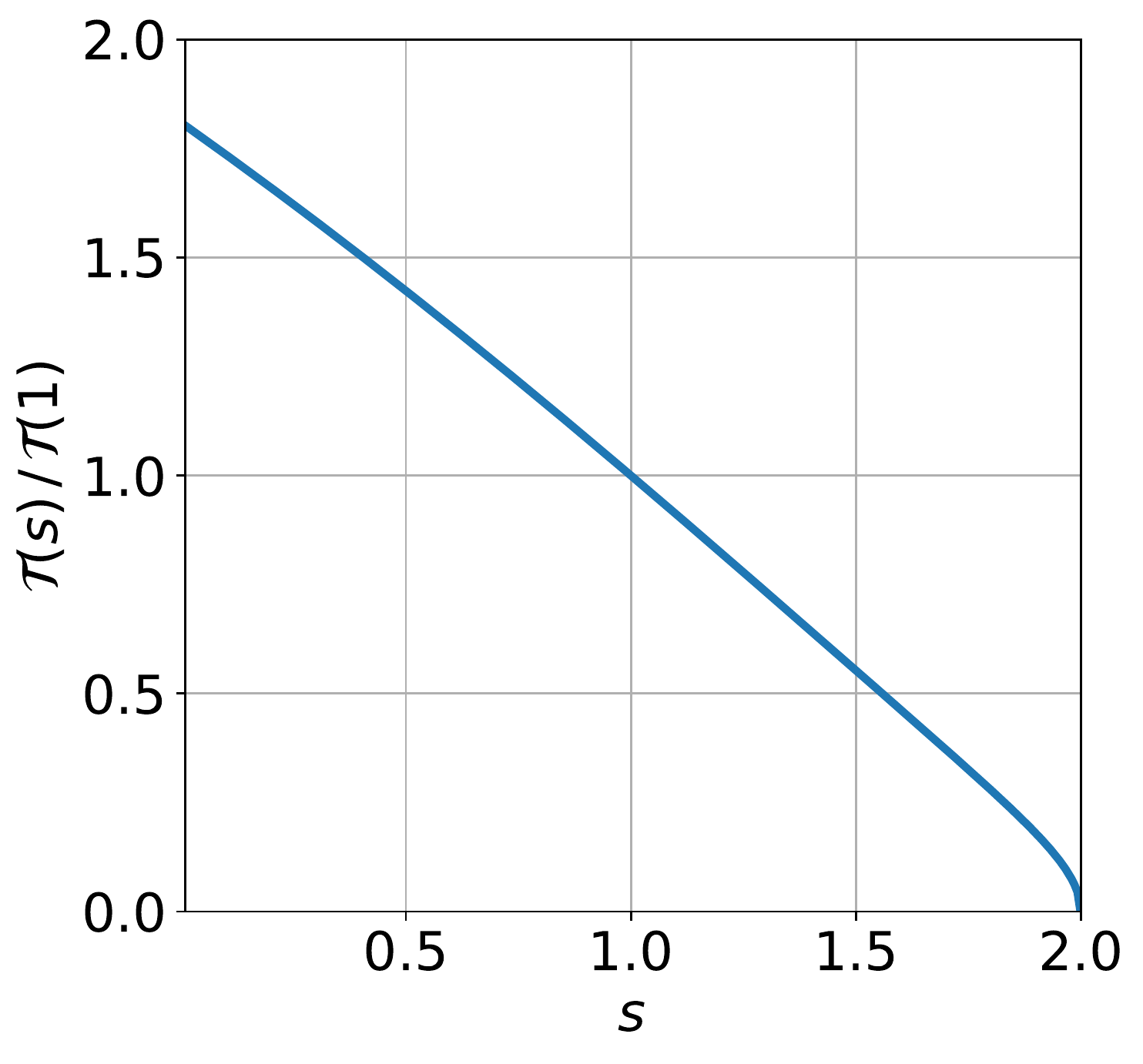}
\caption{Behavior of the function $\mathcal{T}(s)$. The result normalized by the one evaluated at $s=1$  is plotted as a function of the inner slope of the density profile, $s$. 
\label{fig:function_R_s}
}
\end{figure}

In Eq.~(\ref{eq:r_core_general}), the expression of the soliton core size also involves the factor $\tilde{\Psi}(0)+\mathcal{E}/\alpha$, which is given as a function of $\alpha$. To express its dependence on $\alpha$ analytically, we make use of the approximation in the limit of $\alpha\gg1$. As we saw in Eq.~(\ref{eq:alpha_general}), this limit is ensured if we also take the limit $\cvir\to0$ for the inner slope of the halo profile, $s<3$. Then, the key expression is Eq.~(\ref{eq:eigenvalue_semi-analytic}). Similarly to what we did in Sec.~\ref{subsec:Langer_approx},  we use the expression for $\tilde{\Psi}$ valid near the center, i.e., Eq.~(\ref{eq:tilde_Phi_near_origin}). Recalling that the function $g_0$ vanishes at the turning point $\xc\simeq [(3-s)(2-s)\{\tilde{\Psi}(0)+\mathcal{E}/\alpha\}]^{1/(2-s)}$ and assuming $\xc\ll1$, we obtain
\begin{align}
&\alpha^{1/2}\int_0^{x_c}\sqrt{g_0(x')}\,dx'=\Bigl(n-\frac{1}{4}\Bigr)\pi
\nonumber
\\
&\Longrightarrow\,\,
\alpha^{1/2}\frac{\{(3-s)(2-s)\}^{1/(2-s)}}{2-s}\,B\Bigl(\frac{1}{2-s},\,\frac{3}{2}\Bigr)
\nonumber
\\
&\qquad\times\bigl\{\tilde{\Psi}(0)+\mathcal{E}/\alpha\bigr\}^{1/2+1/(2-s)}=\Bigl(n-\frac{1}{4}\Bigr)\pi, 
\end{align}
where the function $B$ is the beta function. 
Setting the integer $n$ to unity (this corresponds to the ground state), the above equation is recast as
\begin{align}
& \tilde{\Psi}(0)+\mathcal{E}/\alpha = \alpha^{-(2-s)/(4-s)}
\Bigl(\frac{3\pi}{4}\Bigr)^{2(2-s)/(4-s)}
\nonumber
\\
&
\times\Biggl[\frac{2-s}{\bigl\{(3-s)(2-s)\bigr\}^{1/(2-s)}}\frac{1}{B(1/(2-s),\,3/2)}\Biggr]^{2(2-s)/(4-s)}. 
\label{eq:energy_eigenvalues_general}
\end{align}

\begin{figure}[tb]
 \includegraphics[width=8.0cm,angle=0]{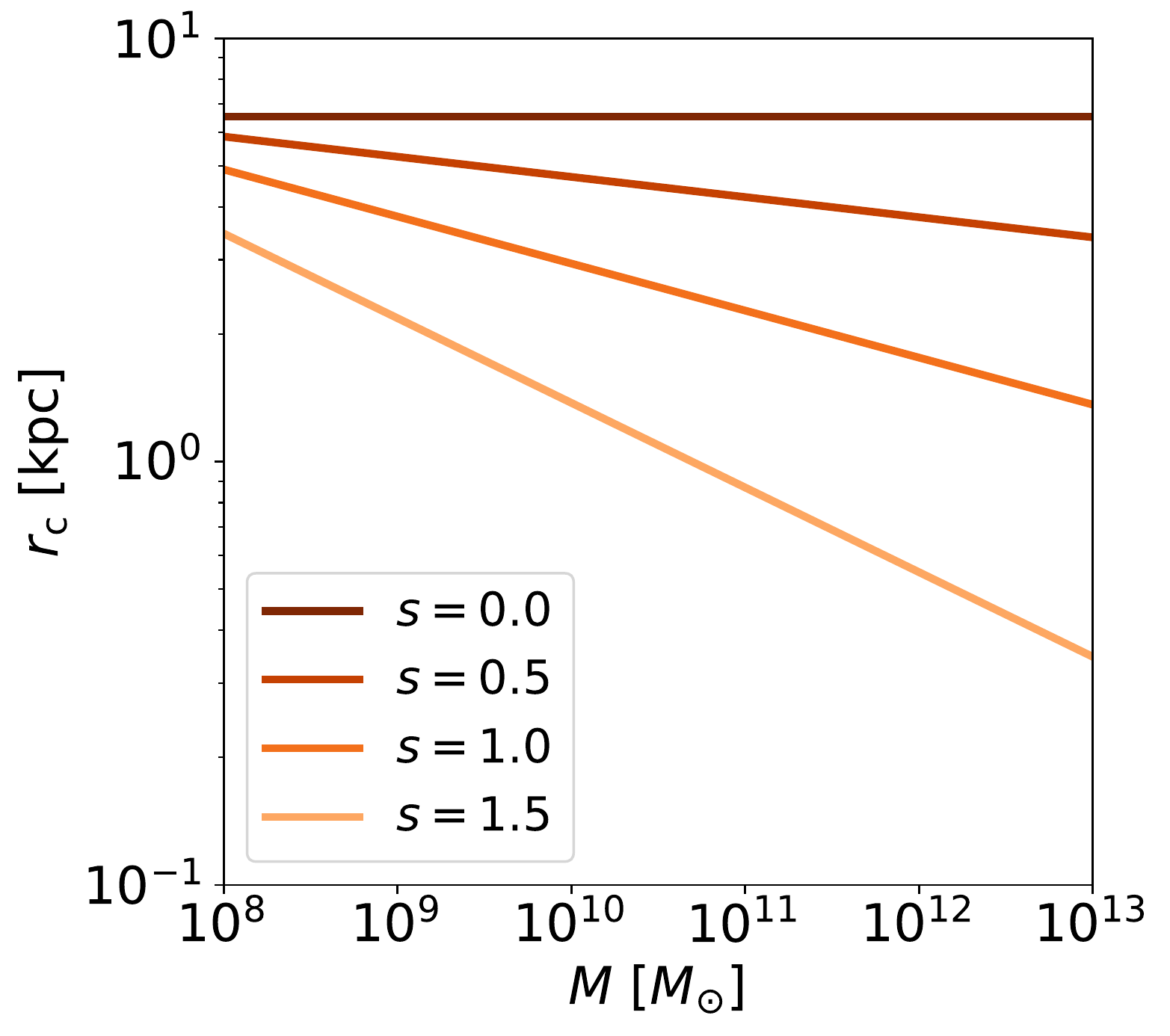}
\caption{Dependence of the core radius-halo mass relation on the inner slope of halo  density profile, $s$. The upper bound on the soliton core radius 
in the limit of $\cvir\to0$, given at Eq.~(\ref{eq:rcore_limit_general}), is plotted as a function of halo mass, fixing model parameters for the cosmology and FDM mass to the default setup (see Sec.~\ref{subsec:core-halo_mass-concentration_relation}). Note that the results for $s=1$ are identical to the lower and upper limits in the core radius and mass shown in Fig.~\ref{fig:core_halo_mass_concentration}.
\label{fig:core-halo_limit_general}
}
\end{figure}

Substituting Eq.~(\ref{eq:energy_eigenvalues_general}) into Eq.~(\ref{eq:r_core_general}), the comoving core radius valid in the limit $\alpha\gg1$ leads to
\begin{align}
\rc&=p\,\frac{6^{1/2}\,\rs}{\alpha^{1/(4-s)}}
\nonumber
\\
&\times
\Biggl[
\frac{4}{3\pi}\,\frac{\bigl\{(3-s)(2-s)\bigr\}^{1/(2-s)}}{2-s}\,B\Bigl(\frac{1}{2-s},\,\frac{3}{2}\Bigr)
\Biggr]^{(2-s)/(4-s)} 
\label{eq:rcore_limit_generalized}
\end{align}
Substituting further Eqs.~(\ref{eq:alpha_general}) and (\ref{eq:r_core_general}) into the above, we arrive at the final expression for the core radius in the limit of $\cvir\to0$:  
\begin{widetext}
\begin{align}
\rc&=3.59\,\,\mbox{[kpc]}\,\,\frac{\mathcal{T}(s)}{\mathcal{T}(1)}
a^{-1/(4-s)}\,\Bigl(\frac{p}{0.65}\Bigr)\Bigl(\frac{m_\phi}{10^{-22}\,\mbox{eV}}\Bigr)^{-2/(4-s)}\Bigl(\frac{M_{\rm h}}{10^9\,M_\odot}\Bigr)^{-s/3/(4-s)}
\Bigl(\frac{\Delta_{\rm vir}}{200}\,\,\frac{\Omega_{\rm m,0}h^2}{0.147}\Bigr)^{-(1-s/3)/(4-s)},
\label{eq:rcore_limit_general}
\end{align}
\end{widetext}
where the function $\mathcal{T}(s)$ is defined by
\begin{align}
& \mathcal{T}(s)\equiv 49.064 \,\bigl\{721.305\,(3-s)\bigr\}^{-1/(4-s)}
\nonumber
\\
&\times
\Biggl[\frac{4}{3\pi}\,\frac{\bigl\{(3-s)(2-s)\bigr\}^{1/(2-s)}}{2-s}\,B\Bigl(\frac{1}{2-s},\,\frac{3}{2}\Bigr)\Biggr]^{(2-s)/(4-s)}.
\label{eq:def_func_T}
\end{align}
Notice that the dependence on the concentration parameter $\cvir$ disappears at Eq.~(\ref{eq:rcore_limit_general}). This means that for a fixed cosmological model and a given halo mass, Eq.~(\ref{eq:rcore_limit_general}) gives a bound on the core radius. Note that setting the slope $s$ to unity, the above expression reproduces the upper bound given at Eq.~(\ref{eq:r_core_upper_limit}). 

Figure~\ref{fig:function_R_s} plots the functions $\mathcal{T}$ normalized by the one evaluated at $s=1$. This shows that the function $\mathcal{T}$ is a monotonically decreasing functions of $s$, and approaches zero at $s\to2$, indicating that for halos with the inner slope of $s=2$, the soliton core disappears. Since the potential at the halo center generally diverges for $s\geq2$, this implies that for $s\geq2$, the fuzzy dark matter soliton, if exists, does not form a flat core, but rather can have a cusp at the center.

Figure~\ref{fig:core-halo_limit_general} shows the core-halo relations in the $\cvir\to0$ limit, focusing specifically on the soliton core radius. 
Setting the model parameters to those adopted in Figs.~\ref{fig:core_halo_mass_concentration}-\ref{fig:core_halo_mass_FDM}, we vary the inner slope of halo profiles from $s=0$ to $1.5$. Here, Steepening the inner slope substantially reduces allowable core sizes, and the predicted curve for $s=1.5$ is inconsistent with simulations.

\bibliographystyle{apsrev4-2}
%


\end{document}